\documentclass[modern]{aastex631}
\NewPageAfterKeywords
\begin{document}

\title{The Nature of a Recently Discovered Wolf-Rayet Binary: \\Archetype of Stripping?}
\altaffiliation{This paper includes data gathered with the 6.5 meter Magellan Telescopes located at Las Campanas Observatory, Chile. It also uses observations made with the NASA/ESA {\it Hubble Space Telescope (HST)}, obtained at the Space Telescope Science Institute, which is operated by the Association of Universities for Research in Astronomy, Inc., under NASA contract NAS 5-26555. These observations were associated with programs GO-16299 (PI: Massey) and GO-16093 (ULLYSES, PI: Roman-Duval).}

\author[0000-0001-6563-7828]{Philip Massey}
\affiliation{Lowell Observatory, 1400 W Mars Hill Road, Flagstaff, AZ 86001, USA}
\affiliation{Department of Astronomy and Planetary Science, Northern Arizona University, Flagstaff, AZ, 86011-6010, USA}
\email{phil.massey@lowell.edu}

\author[0000-0002-5787-138X]{Kathryn F. Neugent}
\affiliation{Lowell Observatory, 1400 W Mars Hill Road, Flagstaff, AZ 86001, USA}

\author[0000-0003-2535-3091]{Nidia I. Morrell}
\affiliation{Las Campanas Observatory, Carnegie Observatories, Casilla 601, La Serena, Chile}

\author{Desmond John Hillier}
\affiliation{Department of Physics and Astronomy \& Pittsburgh Particle Physics, Astrophysics, and Cosmology Center (PITT PACC), University of Pittsburgh, 3941 O'Hara Street, Pittsburgh, PA, 15260, USA}

\author{Laura R. Penny}
\affiliation{Department of Physics and Astronomy, College of Charleston, 208 J.C. Long Building, 9 Liberty Street, Charleston, SC 29424, USA}

\begin{abstract}

LMCe055-1 was recently discovered in a survey for WRs in the Large Magellanic Cloud, and classified as a WN4/O4, a lower excitation version of the WN3/O3 class discovered as part of the same survey. Its absolute magnitude precluded it from being a WN4+O4 binary.  OGLE photometry show shallow primary and secondary eclipses with a 2.2 day period. The spectral characteristics and short period pointed to a possible origin due to binary stripping.  Such stripped WR binaries should be common but have proven elusive to identify conclusively.  In order to establish its nature, we obtained {\it HST} UV and Magellan optical spectra, along with imaging. Our work shows that the WR emission and He\,{\sc ii} absorption arise in one star, and the He\,{\sc i} absorption in another. The He\,{\sc i} contributor is the primary of the 2.2-day system and exhibits $\sim$300~km~s$^{-1}$ radial velocity variations on that time scale.  However, the WR star shows 30-40 km~s$^{-1}$ radial velocity variations, with a likely 35-day period and a highly eccentric orbit.  Possibly LMCe055-1 is a physical triple, but that would require the 2.2-day pair to have been captured by the WR star.  A more likely explanation is that the WR star has an unseen companion in a 35-day orbit  and that the 2.2-day pair is in a longer period orbit about the two.  Such examples of multiple systems are well known among massive stars, such as HD~5980.  Regardless, we argue that it is highly unlikely that the  WR component of the LMCe055-1 system resulted from stripping.

\end{abstract}

\section{Introduction}

One of the greatest uncertainties in massive star research today is the role that binary evolution plays in the evolution of massive stars, and in
particular the formation of Wolf-Rayet (WR) stars.  We know that most WRs are the evolved descendants of massive OB stars.  In the 1960s the prevailing notion was that WRs formed by binary evolution: the outer H-rich layers of a massive star were stripped off by interactions with a close companion, revealing the nuclear processed material underneath \citep{1967AcA....17..355P}.  Initially the CNO-cycle H-burning equilibrium products (nitrogen and helium) would surface, and a WN-type WR would result.  Further stripping would reveal the products of He-burning (carbon and oxygen), and the star would be spectroscopically identified as a WC-type WR. Then came the discovery that mass-loss via radiatively driven stellar winds was ubiquitous among massive OB stars \citep{1967ApJ...147.1017M}, leading \citet{1975MSRSL...9..193C} to propose that this stripping took place not through binary interactions but via stellar winds.   This ``Conti scenario" became the accepted explanation for the formation of WRs for the next 30 years. However, the paradigm has been shifting back towards the binary explanation: mass-loss rates on the main-sequence are lower than we once thought \citep{2006ApJ...637.1025F},  and estimates of binary parameters suggest that many massive stars may interact  during their lifetimes (e.g., \citealt{sana12}).

However, if current binary models are correct, we would expect to find many examples of these stripped binaries.  (For recent reviews, see \citealt{2018ApJ...867..125D} and \citealt{2018A&A...615A..78G}.)  These stripped objects should be very hot, as they are basically bare stellar cores, and long-lived.   Spectroscopically, the lower-mass versions should resemble subdwarf O (sdO) stars. Several sdO stars with Be-type companions have been found and described as stripped binaries (\citealt{2017ApJ...843...60W,2018ApJ...853..156W,2021AJ....161..248W} and references therein).  \citet{2023ApJ...959..125G} have analyzed a sample of two dozen stripped stars of intermediate mass; their spectra are basically that of O-type stars but with little hydrogen (i.e., that of a classic sdO).   But the higher mass versions should resemble WR stars \citep{2018ApJ...867..125D,2018A&A...615A..78G}. Where are they?  We should expect young
clusters such as h and $\chi$ Per (age 12~Myr) to be full of such objects, yet none are found.  Is this a selection effect, with such stars
swamped by the light of a now visually brighter companion, as suggested by \citet{2014ApJ...782....7D}? Recent binary modeling by \citet{2018A&A...615A..78G} has shown that such objects should have WR-like emission lines, ``semi-transparent atmospheres" (and hence absorption lines), small radii (0.2-1$R_\odot$), and high temperatures (20,000-100,000~K).    This turns out to be an excellent description of a newly discovered WR binary in the LMC, LMCe055-1 (OGLE-LMC-ECL-03548), the subject of the present paper.

\clearpage

\subsection{The WN3/O3 Class and the One That Did Not Belong}

\citet{FinalCensus} describe a four-year long survey for WRs  in the Magellanic Clouds.
Among their other discoveries, they found a new class of WRs, which they called WN3/O3s.  They have the emission-line characteristics of high-excitation WN stars (WN3s) but absorption spectra typical of the hottest O-type stars, O3~V.   These stars were far too faint, however, to be WN3+O3 binaries, as their absolute magnitudes were about $-2.5$ to $-3$, while an O3~V star by itself would have $M_V\sim-6$. Altogether, nine of these stars were found \citep{FinalCensus}, which represent about 8\% of the LMC's WN population: they are not rare, one-of-a-kind objects.   \citet{NeugentWN3O3s} analyzed ground-based optical and {\it HST} UV spectra of these stars, finding that they could reproduce the emission and absorption with a single set of physical parameters.  The WN3/O3 properties are similar to other high-excitation WNs in the LMC, except that their mass-loss rates are a factor of 3-5 times lower, more like that of an O-type star.  \citet{NeugentWN3O3s,FinalCensus} discuss possible origins for this class, including the possibility that they are stripped binaries, but there are several arguments against this, including that their radii and masses are larger than that predicted by binary evolution models.  None show signs of binarity, either via radial velocity variations or from eclipses \citep{2023ApJ...947...77M}.  Their spatial distribution in the LMC is indistinguishable from that of the other high excitation WNs, suggesting a common origin \citep{FinalCensus}. 

However, there was a tenth star, LMCe055-1 (discovered  late in the survey), that was not quite like the others, and does not fit neatly into the WN3/O3 class \citep{MasseyMCWRIII}.
Spectroscopically it is of lower excitation, both in terms of the emission and absorption, more like a WN4/O4.  It had previously been identified as a 2.2-day eclipsing binary by the Optical Gravitational Lensing Experiment (OGLE); its light curve showed shallow ($\le0.1$ mag) primary and secondary eclipses \citep{2003AcA....53....1W,2011AcA....61..103G}.  
The star is located  at $\alpha_{\rm 2000}$=04:56:48.81, $\delta_{\rm 2000}$=-69:36:40.6 within the Lucke-Hodge 8 ``star cloud" \citep{LuckeHodge}, with a $V$ magnitude of 16.2.  Its Gaia DR3 proper motion and parallax \citep{gaia,gaiadr3} demonstrate that is a member of the LMC.   In this paper we describe our observational campaign, and accompanying challenges,  to understand this star and determine if it is perhaps the archetype of a high-mass stripped binary. 

In Section~\ref{Sec-obs} we describe our observations; in Section~\ref{Sec-analysis} we present our corresponding analysis, including our spectral modeling and radial velocity measurements.  In Section~\ref{Sec-system} we tie these together to arrive at a description of the LMCe055-1 system. We summarize our conclusions in Section~\ref{Sec-conclude}.

\section{Observations and Reductions}
\label{Sec-obs}

Our observations were aimed at achieving three goals.  First, we wanted to model the physical properties of the star (or at least that of the primary) using the same techniques \citet{NeugentWN3O3s} had employed for the WN3/O3s. Second, we wanted to identify and classify the companion star involved in the eclipses.  And third, we wanted to obtain radial velocities for the primary (and hopefully the secondary) in order to determine an orbit and hence masses.  In order to achieve all this, we obtained time-series optical spectra,  UV spectra, and V-band photometry. This section describes these data and their reductions.  

\subsection{Spectroscopy}
\label{Sec-spectra}
\subsubsection{Optical}

The spectrum which led to our classification of LMCe055-1 as a ``WN4/O4" object was obtained on UT 2016 January 11, as described by \citet{MasseyMCWRIII}.  It, and all but two of the subsequent spectra, were obtained with the Magellan Echellette Spectrograph (MagE) mounted on the folded port of the Baade 6.5 m Magellan telescope.  The instrument design is discussed in detail by \citet{MagE}; see also the description given in the user manual.\footnote{https://www.lco.cl/technical-documentation/the-mage-spectrograph-user-manual/}   We used a 1\farcs0 $\times$ 10\arcsec\ slit, yielding a spectral resolving power $R\sim$4100. Wavelength coverage was from 3100~\AA\ to 1~$\mu$m distributed over echellette orders 16 through 6.  Operation is simple, with the target centered on the slit by the telescope operator using a slit-viewing camera.  The slit was always oriented to the parallactic angle. 

Our 2016 discovery spectrum consisted of only a single 700~s exposure, but subsequent observations (obtained through 2022) aimed for higher signal-to-noise (S/N),  generally employing $3\times$15 min or $3\times$20 min exposures, depending upon the seeing. The multiple exposures were used to eliminate radiation events (``cosmic rays").  Details of our MagE flat-fielding and reduction procedures can be found in \citet{2012ApJ...748...96M}.   Table~\ref{tab:spectra} lists our spectroscopic observations.  Once we had the light-curve ephemeris well established (see Section~\ref{Sec-epherm}), we generally timed our observations to phases near quadrature (i.e., phases 0.25 and 0.75, where phase 0 corresponds to the time of primary eclipse), as this is when we expected the separation between primary and secondary stars would be the largest.  As we will see, the actual situation was far more complicated than that.

\begin{deluxetable}{l c l r r c c l}
\tabletypesize{\scriptsize}
\tablecaption{\label{tab:spectra}Optical Spectroscopic Observations LMCe055-1}
\tablewidth{0pt}
\tablehead{
\colhead{HJD-2450000}
&\colhead{Phase\tablenotemark{a}}
&\colhead{UT Date}
&\colhead{S/N\tablenotemark{b}}
&\colhead{Exp.\ time (s)}
&\colhead{Seeing (\arcsec)}
&\colhead{Instrument}
&\colhead{Comment} 
}
\startdata
 7398.607 & 0.117 & 2016 Jan 11 & 50 & 1$\times$700 & 0.6 & MagE &  \\
 7791.549 & 0.115 & 2017 Feb 7 &  35 & 3$\times$900  & 1.6& MagE &ThAr lost, low S/N, Not used for RVs\\
 7791.696 & 0.183 & 2017 Feb 7 &  30 & 3$\times$550  & 1.0 & MagE & Low S/N, Not used for RVs \\
 7792.573 & 0.589 & 2017 Feb 8 & 95 & 3$\times$1200  & 0.6 & MagE & \\
 7792.698 & 0.647 & 2017 Feb 8 & 63 & 3$\times$900  & 0.9 & MagE &  \\
 7793.691 & 0.107 & 2017 Feb 9 & 60 & 3$\times$900 & 0.8 & MagE & \\
 8118.552 & 0.572 & 2017 Dec 31 & 70 & 3$\times$900& 0.9 & MagE & \\
 8119.553 & 0.036 & 2018 Jan 1 & 100 & 3$\times$900  & 0.7 & MagE &  \\
 8124.552 & 0.351 & 2018 Jan 6 & 90 & 3$\times$900 & 0.7 & MagE & \\
 8153.541 & 0.778 & 2018 Feb 4 & 100 & 3$\times$900 & 0.7& MagE & Independent\tablenotemark{c} \\
 8153.574 & 0.793 & 2018 Feb 4 & 100 & 3$\times$900 & 0.6& MagE & Independent\tablenotemark{c} \\
 8153.556 & 0.785 & 2018 Feb 4 & 130 & 6$\times$900 & 0.6-0.7 & MagE & Combined\tablenotemark{c} \\
 8154.577 & 0.258 & 2018 Feb 5 & 100 & 3$\times$1200 & \tablenotemark{c} & MagE & Independent\tablenotemark{c} \\
 8154.621 & 0.278 & 2018 Feb 5 & 100 & 3$\times$1200 & \tablenotemark{c} & MagE & Independent\tablenotemark{c} \\
 8154.598 & 0.268 & 2018 Feb 5 & 135 & 6$\times$1200 & \tablenotemark{c} & MagE & Combined\tablenotemark{c} \\
 8440.641 & 0.754 &2018 Nov 18 &125& 3$\times$1200 & 0.6 & MagE & \\
 8482.748 & 0.256 & 2018 Dec 30 &170 & 3$\times$1200 & 0.6 & MagE & \\
 8864.789 & 0.205 & 2020 Jan 16 & 150 & 3$\times$1200 & 1.0 & MagE & \\
 9187.668 & 0.753 & 2020 Dec 4  & 100 & 3$\times$1200 & \tablenotemark{d} & MagE & \\
 9253.557 & 0.270 & 2021 Feb 8  & 130 & 5$\times$1000 & $<$0.9 & GMOS \\
 9254.585 & 0.746 & 2021 Feb 9  & 130 & 5$\times$1000 & $<$0.9 & GMOS \\
 9854.829 & 0.760 & 2022 Oct 2  & 110 & 3$\times$1200 & 0.7 & MagE & Lost 150 s due to tel.\ runaway\\
  \enddata
 \tablenotetext{a}{Based upon our final eclipse ephemeris given in Section~\ref{Sec-epherm}, with T0=2457001.0906 and P=2.159044~d.}
 \tablenotetext{b}{Signal-to-noise ratio per $\sim$0.25~\AA\ pixel in line-free region, roughly 4400-4425~\AA.}
 \tablenotetext{c}{The observations obtained on JDs 2458153 and 2458154 consisted of two consecutive sets of 3$\times$1200s exposures, each reduced independently with their own comparison arcs in order to better evaluate our measuring uncertainty.  The two sets were then averaged to produce one combined spectrum which was used in our analysis.}
 \tablenotetext{d}{Seeing data not available.}
 \end{deluxetable}

Two additional observations were obtained with the Gemini Multi-Object Spectrograph (GMOS) mounted on the Gemini-South 8 m telescope. The observations were made possible thanks to the Gemini ``fast-turn around" program, and were obtained on two successive nights, (UT) 2021 February 8 and 9 under program GS-2021A-FT-103.  (The fact that these were taken on two successive nights proved very helpful in our interpretation, as discussed in Section~\ref{Sec-system}.)  The observations were taken at phases 0.27 and 0.75.  Each night's data consisted of 5 consecutive exposures of 1000s each with the B1200 grating centered at 4550~\AA\ using a 0\farcs5 wide slit   The dispersion was 0.26~\AA\ pixel$^{-1}$, and the resulting spectral resolving power was $R\sim3750$, similar to that of our MagE data.  The detector consists of three  Hamamatsu CCDs, each with 2048$\times$4176 15$\mu$m pixels, arranged in a row to provide 6266 pixels in the dispersion direction. (See \citealt{2016SPIE.9908E..2SG} and the GMOS manual.\footnote{https://www.gemini.edu/instrumentation/gmos/})  The data were binned by two pixels in the spatial direction. Wavelength coverage on the three chips were 3740-4255~\AA, 4281-4799~\AA, and 4827-5351~\AA, with the missing wavelengths due to the gaps between the chips.  An arc (CuAr) and internal flat (``GCALflat") preceded each set of the five LMCe055-1 exposures.  

The GMOS data were reduced using the Gemini reduction routines in {\sc iraf}\footnote{NOIRLab IRAF is distributed by the Community Science and Data Center at NSF NOIRLab, which is managed by the Association of Universities for Research in Astronomy (AURA) under a cooperative agreement with the U.S. National Science Foundation.}  \citep{1986SPIE..627..733T,1993ASPC...52..173T,2024arXiv240101982F}.  During the reductions, we found that the data on the reddest section (the Gemini documentation refers to it as ``CCD-1") was extremely noisy, with many spikes. The Gemini Observatory had previously informed us that there was a serious intermittent charge-transfer-efficiency problem with this device, and that our data had been taken during one of the affected periods.  Despite kind help from Dr.\ Vinicius Placco at NOIRLab, we were unable to do anything that could make data from that chip useful, and so we restricted our analysis to the two unaffected chips.

Our MagE data were flux calibrated using observations of multiple spectrophotometric standards made each night.  A comparison of our spectra showed that in general the shapes of the flux distributions are well determined, to 3\% or better.  However, the zero-point differs from observation to observation.  This is not due to the small light variability (after all, most spectra were taken at similar phases and the eclipse depths are small) but rather due to the differences in slit losses between the standards and target observations.  Since a comparison between our model and observed fluxes will be of interest, we scaled the spectra used for modeling following a similar procedure to that employed by \citet{ErinWC,ErinWO}.  For comparison with our modeling we selected our spectrum with the best S/N, the MagE spectrum taken on 2018 December 30.  We convolved its fluxed spectrum $F_\lambda$ with the V-band sensitivity curve $S_\lambda$ of \citet{1990PASP..102.1181B}.  
The derived V-band magnitude from our spectrum then is 
$$ -2.5 \log (\frac{\int F_\lambda S_\lambda \,d \lambda}{\int S_\lambda \,d \lambda})  -21.10,$$
where the constant comes from Appendix A in \citet{1998A&A...333..231B}.  For our fluxed MagE spectrum, we derived an instrumental V magnitude of 16.07.  Our photometry, presented in the next section, suggests the true V magnitude is 16.17, and so we have corrected our fluxed spectrum by a factor of 0.91.

No attempt was made to flux calibrate the GMOS data.

\subsubsection{Ultraviolet}

For modeling hot stars, spectrophotometry in the ultraviolet (UV) is extremely useful.  Combined with the optical, it provides a longer baseline for determining the correction for interstellar reddening; it gives us useful data on the flux close to the peak of the spectral energy distribution.
Most importantly, it provides us with critical resonance and subordinate spectral lines that provide crucial information on the terminal velocities and strong constraints on the mass-loss rates, abundances, temperatures, and other stellar parameters.  (For a recent review, see \citealt{2020Galax...8...60H}.)

We proposed several times to obtain UV spectra of LMCe055-1 with the Cosmic Origins Spectrograph (COS) on the Hubble Space Telescope (HST), finally succeeding in making our case in Cycle 28 under program ID 16299 (PI: Phil Massey).  A  5059 s exposure  through the 2\farcs5 Primary Science Aperture was obtained with the G140L/800 setup on 2020 Oct 3; the dataset is LEA401010. This provided wavelength coverage 920-1950~\AA\ with a dispersion of 0.08~\AA\ pixel$^{-1}$.  The spectral resolution is 7.8 pixels (0.55~\AA) and thus we were comfortable smoothing the data using a boxcar average of 7 pixels in order to decrease the noise.  The resulting spectral resolving power varies from $R\sim$1700 (175 km~s$^{-1}$) at the short wavelength end to 3500 (85 km~s$^{-1}$) at the long wavelength end.   We used the fully calibrated pipeline reduction of the data.  The S/N is about 60 per 0.55~\AA\ spectral resolution element at 1500~\AA. 

Ironically, LMCe055-1 proved to be a popular target for UV spectroscopy in this time period. First, unbeknownst to us at the time we prepared our Cycle 28 proposal, the object had been scheduled to be observed in Cycle 27 under program ID 15824 (PI: Nathan Smith) along with others of our recently discovered WN3/O3 stars.  It was successfully observed on 2020 Jun 22, just four months before our Cycle 28 program executed.  Second, LMCe055-1 had  been an add-on target to the Director Discretionary program ULLYSES (UV Legacy Library of Young Stars as Essential Standards) observing list.  Although the focus of the ULLYSES project is on normal OB stars, community input advocated for including some short-period binaries as well, and our discovery description of LMCe055-1 \citep{MasseyMCWRIII} had apparently generated some interest, particularly among the stripped binary enthusiasts.   The first ULLYSES observation of LMCe055-1 failed due to guide star acquisition failure, but subsequent exposures were successfully obtained on UT 2021 Dec 5, fourteen months after our observation.  Those data were taken under program ID 16093 (PI: Julia Roman-Duval). 

The ULLYSES data consist of a 2060 s exposure with the G130M/1291 setting (LEHDFC010) and a 4700 s exposure with the G160M/1611 setting (LEHDFC020).  Together, these spectra covered 1150~\AA\ to 1780~\AA\ at a spectral resolution of $\sim$0.09~\AA, corresponding roughly to 8-9 pixels.  Degraded to the same spectral resolution as our G140L exposures, the S/N is quite similar.  Our G140L spectrum has greater wavelength coverage, but the ULLYSES spectrum has greater spectral resolution, so the two are complementary. We were fortunate to obtain archival support (HST-AR-17553) for including these data in our study here.  

Unfortunately, the ID 15824/Smith UV spectrum did not prove similarly useful.  Like one of the ULLYSES exposures, it was obtained with the G140M/1611 setting and covered the 1360-1775~\AA\ region, but since the exposure time was much shorter (2200 s vs.\ 4700 s) the S/N is correspondingly lower. We do not include that in our study, other than to note that it, like the other three UV spectra, all agree well in terms of the absolute flux. 

\begin{figure}[ht!]
\epsscale{0.80}

\vskip -130pt
\plotone{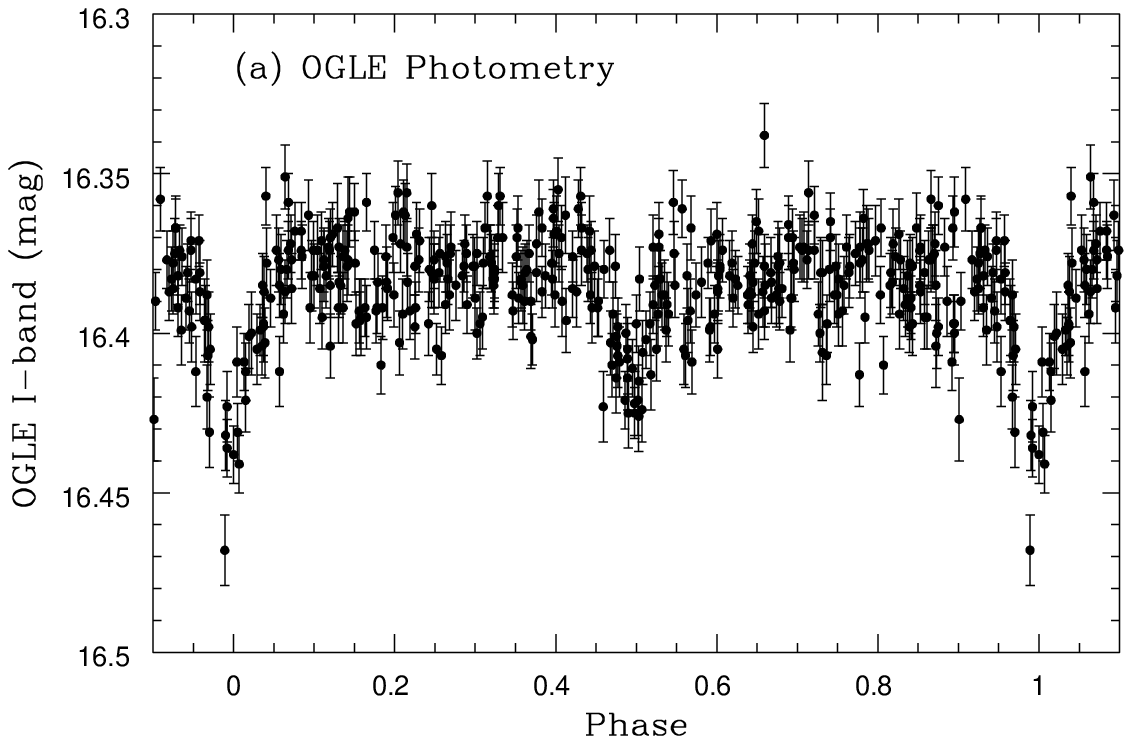}

\vskip -100pt
\plotone{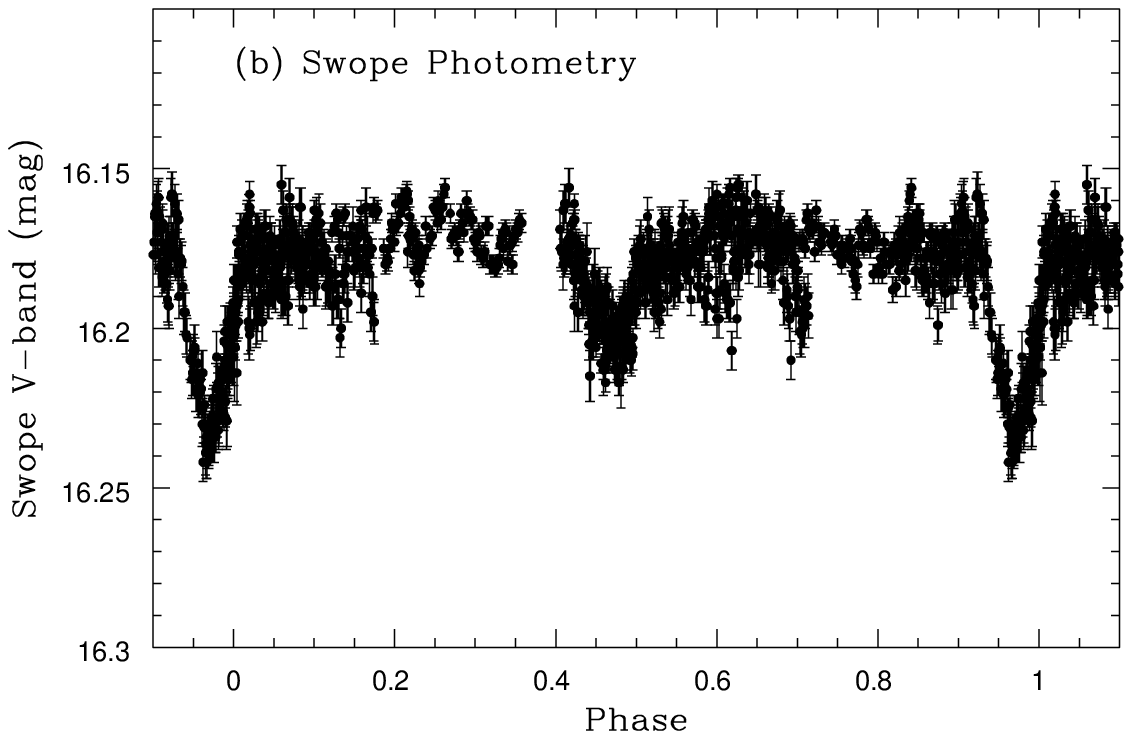}

\vskip -100pt
\plotone{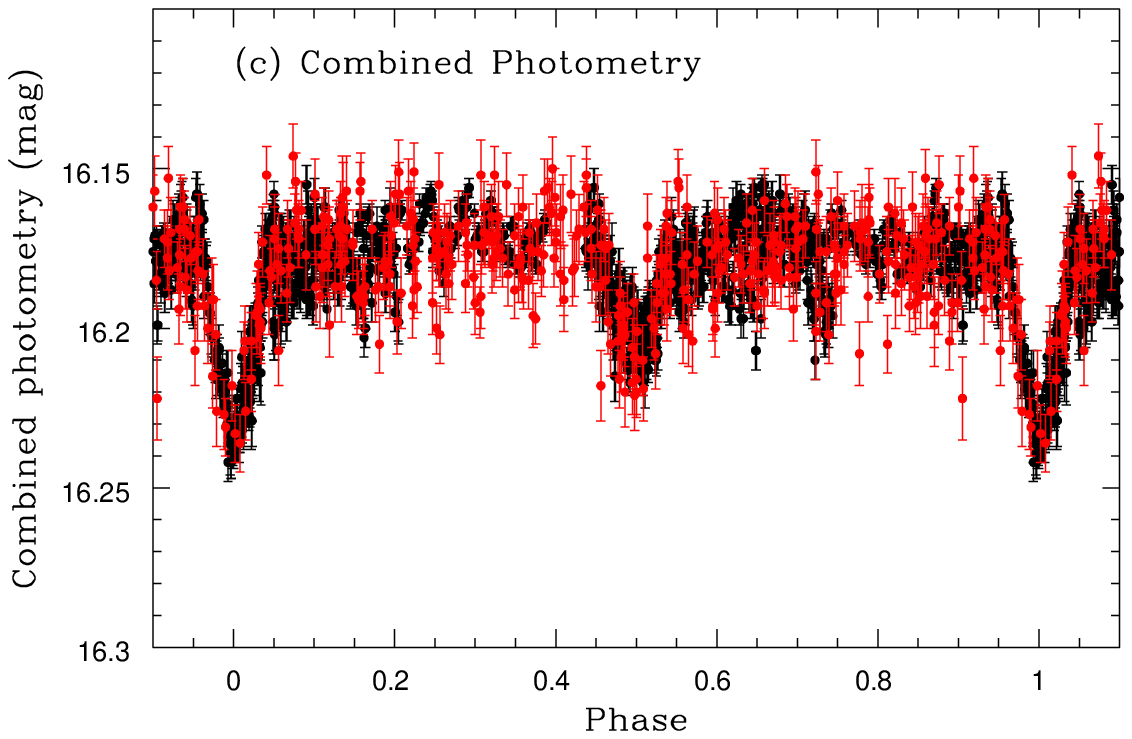}
\caption{Phased photometry.   (a) The OGLE I-band photometry is phased with the OGLE ephemeris of T0=2457001.142  and P=2.1590747.  Although the data are noisy, they clearly reveal both a primary and secondary eclipse. (b) Phasing the Swope data with the OGLE ephemeris reveals a small phase shift, as evidenced by the eclipses not quite lining up with phases of 0 and 0.5. (c)
The data (Swope, black; OGLE, red) have now been phased using our final adopted ephemeris T0=2457001.0906 and P=2.159044~d. The OGLE I-band data have been shifted by -0.205~mag in order to overlay on the Swope V-band data.
\label{fig:phot}}
\end{figure}

\subsection{Imaging}
\label{Sec-phot}
LMCe055-1 was identified as an eclipsing binary by OGLE  in their Phase III
experiment, which was carried out during 2001-2009, using the dedicated 1.3 m Warsaw telescope on Las Campanas \citep{1997AcA....47..319U,2005AcA....55...43S}.
It was cataloged as OGLE LMC-ECL-3548  with a period of 2.1590747 d, and a time of primary
eclipse T0 of 2457001.142 \citep{2011AcA....61..103G}, based on 441 I-band measurements that ranged in time from
2452166.9 (2001 September 14) to 2454947.5 (2009 April 26).  As shown in Figure~\ref{fig:phot}a, the data revealed a primary eclipse with a depth of about $\sim$0.07 mag and a secondary eclipse of $\sim$0.05 mag; individual points typically had photometric uncertainties of $\sim$0.01~mag.   Given the shallowness of the eclipses compared to the OGLE photometric errors, and the relatively sparse coverage during the eclipses, we initiated a campaign to obtain our own data.

\subsubsection{Data Taking and Basic Reductions}
Our imaging observations were all carried out using the f/7 Henrietta Swope  1~m telescope on Las Campanas, using a 4K$\times$4K e2v CCD camera read out through four amplifiers.  The field around LMCe055-1 is not particularly crowded, and the image scale of 0\farcs435 pixel$^{-1}$ was well suited to our project.  The large field of view (29\farcm7$\times$29\farcm8)  allowed an excellent selection of comparison stars for our differential photometry.  Care was taken to slightly offset the telescope 60\arcsec-90\arcsec to the north-east so that LMCe055-1 would always be read out through the same amplifier (designated ``CCD-3" although this is simply the third quadrant of a single CCD). 

\begin{figure}
\plotone{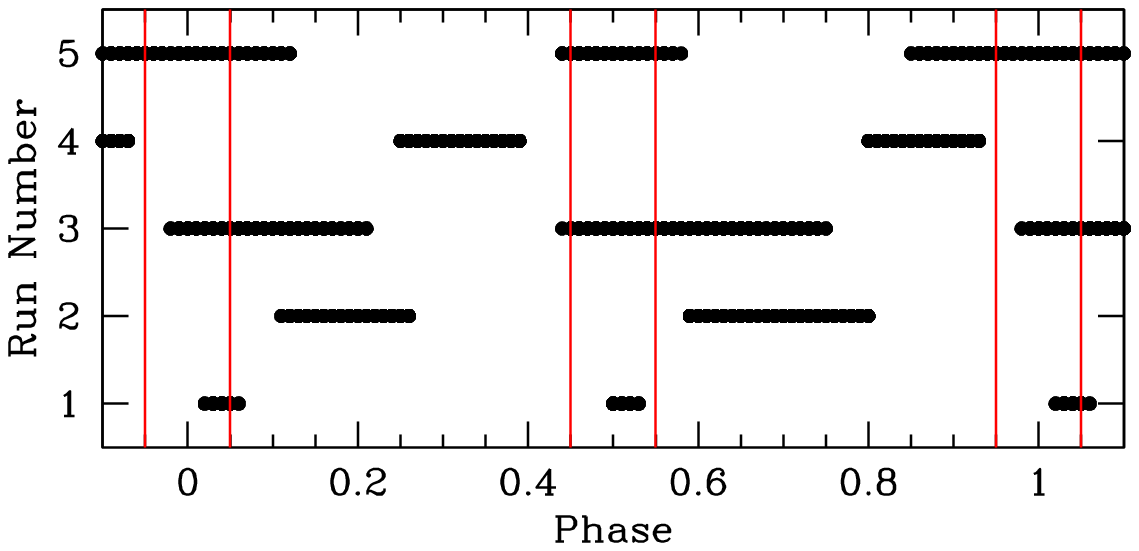}
\caption{\label{fig:coverage} Phase coverage for each of the five imaging runs.  Note that primary eclipse occurs at a phase of 0.0 and secondary eclipse. at a phase of 0.5.  In each case the eclipse extends roughly $\pm$0.05 in phase, as shown by the red lines. For clarity, the coverage near primary eclipse is shown at a phase of both 0 and 1.  Despite the large amount of data and the relatively short period, there is still a small gap in phase coverage around a phase of 0.4.}
\end{figure}

Table~\ref{tab:Swoperuns} lists the dates of our observing runs.  All observations were made through the V filter.  Our typical observing procedure was to obtain twilight flats, set up on the field, initiate the guider, and then begin a continuous series of 150 s exposures for the rest of the night.
The readout time is 37 s, and so in a typical night we could obtain well over 100 observations.  The exception to this procedure was on the first set of data, which were kindly obtained by Dr.\ George Jacoby during short gaps in his own observing program, timed to coincide with primary and secondary eclipses.   Observations obtained on the first night of (10 November 2017) that run were 120 s long; observations on the second night (11 November 2017) were 150 s in length.   We include in Table~\ref{tab:Swoperuns} the median photometric error of our photometry; on runs when the moon was bright and nearby, the photometric precision is unsurprisingly lower.   The phase coverage for each of the five runs is shown in Figure~\ref{fig:coverage}.

All observing nights were photometric.  Although the seeing at Las Campanas is world-renowned, optical aberrations in the Swope optics generally limit the delivered image quality (DIQ). As shown in Table~\ref{tab:Swoperuns}, stars on our images typically had full-width-at-half-maximum values of 1\farcs6 (3.7 pixels).

\begin{deluxetable}{c l r c c c l}
\tabletypesize{\scriptsize}
\tablecaption{\label{tab:Swoperuns}Swope Photometry Runs}
\tablewidth{0pt}
\tablehead{
\colhead{Run}
&\colhead{UT Dates}
&\colhead{\# Obs.}
&\colhead{Med.\ Err}
&\multicolumn{2}{c}{90\% Seeing}
&\colhead{Comment} \\ \cline{5-6}
& & & (mag) &  \colhead{Med} & \colhead{Range}  
}
\startdata
1        &2017 Nov 10-11  & 74 & 0.003 & 1\farcs6 &1\farcs3-1\farcs8 & Partial nights\\
2        &2017 Nov 18-20  &358 & 0.003 & 1\farcs6 &1\farcs5-2\farcs0 & \\
3        &2017 Dec 3-7     &728  & 0.005 & 1\farcs8 & 1\farcs6-2\farcs3 & Full moon \\
4        &2017 Dec 25-26    &262 & 0.003& 1\farcs5 &1\farcs3-1\farcs8& \\
5        &2018 Jan 29-Feb 2 &522& 0.006 & 1\farcs6 & 1\farcs4-2\farcs0 & Full moon\\
\enddata
\end{deluxetable}

The data were reduced using our own {\sc iraf} scripts constructed for our previous survey for WRs in Magellanic Clouds; details of the scripts are given in  \citet{MasseyMCWRI} and \citet{MasseyMCWRII}.  After the headers were updated with accurate trim and bias section values, the overscan values were determined and subtracted for each of the four quadrants and the extraneous columns removed.
Each quadrant was then flipped so that north was up and east was to the left.  Small non-linearity corrections were made, and the four quadrants were
stitched together to a single frame.  The resulting images were then 4096$\times$4110 in size. The same process was applied to the individual bias and twilight flats.  The latter were corrected for the iris shutter pattern if the exposures were shorter than 15 s in length.  The bias frames were averaged, and subtracted from each of the flats and program frames.  The twilight flats were scaled so they have a common mode and then combined with an algorithm that rejects any points that are 3.5$\sigma$ more than expected on the basis of photon- and read-noise, thus eliminating any stars
or cosmic rays in creating the master flat.  The normalized flat was then divided into each of the program frames.   A local version of the ``astrometry.net" software \citep{lang} was then used to revise the world-coordinate-system of the program images using 2MASS \citep{2MASS} coordinates as the reference frame.

\subsubsection{Doing the Photometry}
  We performed aperture photometry using our own {\sc iraf} and {\sc fortran} pipeline, which automatically determined the DIQ on each image, and performed star-finding and photometry on all the stars on each frame.  These routines relied upon using the standard {\sc iraf} ``daofind" and ``phot" routines, employing aperture radii of 3, 5, and 7 pixels, with sky determined from the modal values from an annulus 10-20 pixels from each object. 

With nearly 2000 observations obtained during 17 nights, we were able to experiment with the best way to produce an accurate light-curve for LMCe055-1.   We began by identifying isolated stars (no neighbors within 10 pixels) from the photometry of a single reference frame, choosing stars that were in common to the calibrated V-band photometry of \citet{ZaritskyLMC} and which had photometric errors of 0.01~mag or less.  
This produced a set of 759 potential reference stars.

We tested three subsets of these reference stars. One was the complete set of 759 stars.  Another was a subset that had fallen on the same subframe (CCD-3) as  LMCe055-1 (167 stars).  The third was simply a subset of the reference stars within 300\arcsec\ of LMCe055-1 (89 stars).   We constructed nine zero-points for our reference photometry, one for each combination of the three apertures and subsets.  

We then tested to see which of these nine samples worked the best.  We identified 38 stars near LMCe055-1 that were of similar brightness, and tested to see which of our apertures and reference subsets gave the lowest scatter frame-to-frame for these 38 stars.  (Of course, we allowed for the possibility that some of these might be variable, and were careful to ignore outliers.) The best results were obtained for the photometry obtained with a radius of 5 pixels, which is what we expected from first principles given the typical seeing (fwhm's of 3-4 pixels).\footnote{See discussion about optimal aperture size in https://iraf.net/irafdocs/daophot2/.} The subset of 89 reference stars within 300\arcsec of LMCe055-1 gave marginally smaller scatter than the other two reference sets, and we adopted it.  These tests confirmed that our results were robust and all achieved the desired millimag precision. 
The final photometry is given in Table~\ref{tab:phot} and has been smoothed by a three-point running median filter. Given that the observations
are typically spaced 187 s apart, and that the period is 2.2 days, the maximum smearing is 0.003 orbital phase, which is inconsequential.

\begin{deluxetable}{l c c c}
\tabletypesize{\scriptsize}
\tablecaption{\label{tab:phot}Swope Photometry}
\tablewidth{0pt}
\tablehead{
\colhead{HJD-2450000}
&\colhead{V}
&\colhead{Verr}
&\colhead{Run}
}
\startdata
8067.6984   & 16.213 &   0.004& 1 \\  
8067.7012   & 16.228   & 0.003  & 1 \\
8067.7031   & 16.213  & 0.004  & 1 \\
8067.7049   & 16.205  &  0.003  & 1 \\
8067.7068   & 16.205   & 0.004  & 1 \\
8067.7086   & 16.210    &0.004  & 1 \\
8067.7104   & 16.212&    0.003  & 1 \\
8067.7123   & 16.212 &   0.004  & 1 \\
8067.7141   & 16.201&   0.003  & 1 \\
8067.7160   & 16.201 &   0.003  & 1 \\
8067.7197  &  16.202 &   0.004  & 1 \\
8067.7215    &16.202  &  0.003  & 1 \\
\enddata
\tablecomments{Table~\ref{tab:phot} is published in its entirety in the machine-readable format.
      A portion is shown here for guidance regarding its form and content.}
\end{deluxetable}

\clearpage
\subsubsection{Refinement of the Eclipse Ephemeris}
\label{Sec-epherm}

In order to refine the ephemeris, we
began by performing our own period search on the OGLE I-band photometry using Dr.\ Marc Buie's IDL implementation\footnote{https://www.boulder.swri.edu/$\sim$buie/idl/} of the phase dispersion minimization (pdm) technique described by \citet{1978ApJ...224..953S}. 
 We found a slightly shorter period than OGLE had found,  P=2.159067~d (rather than 2.159075~d); a similar result (P=2.159063~d) was found if the two apparently anomalous data points (one high, one low) apparent in the OGLE data (Figure~\ref{fig:phot}a) were removed.  

Phasing our Swope data (taken in late 2017 to early 2018) to the OGLE ephemeris (September 2001-April 2009) showed a shift of 0.035  cycles for the eclipses (Figure~\ref{fig:phot}b).  We therefore set about refining the ephemeris using the combined data.  The Swope data are much more finely sampled, with 1944 V-band measurements obtained over a 3 month period as discussed above.  The depths of the eclipses are similar in both the OGLE I-band photometry and
the Swope V-band photometry, and we combined the data sets by applying a shift of -0.205~mag to
the OGLE data. However, the vast difference in sampling frequency and the large gap between the two data sets proved challenging
for the pdm technique.  We therefore performed a period search using the Lafler-Kinman technique \citep{1965ApJS...11..216L}, a brute-force approach which phases the data by trial periods and calculates the point-to-point
scatter, which will be a minimum when the data are correctly phased for well sampled light-curves.  The Lafler-Kinman search  yielded a best period of 2.159045 d.  

As a further check, we refined the period using a third method.  For each of the two data sets (OGLE and Swope) we then computed the phase of maximum eclipse using a small range of trial periods. The best period should then correspond to each data set having the same phase for minimum light. (As long as the same
value for T0 was adopted for all the trials, its value could be arbitrary.) The value found by this matter
was P=2.159044~d, in near-perfect agreement with the 2.159045~d period found above using the Lafler-Kinman search on the combined data set.  We determined an accurate value for T0 as part of the light-curve fitting described in Section~\ref{Sec-2.2bin}, adopting 2457001.0906.  We adopt this and the 2.159044~d period as our final ephemeris.  The agreement in the phasing between the two data sets is quite satisfying, as shown in  Figure~\ref{fig:phot}c.

\section{Analysis}
\label{Sec-analysis}

\begin{figure}
\epsscale{1.2}
\vskip -150pt
\plotone{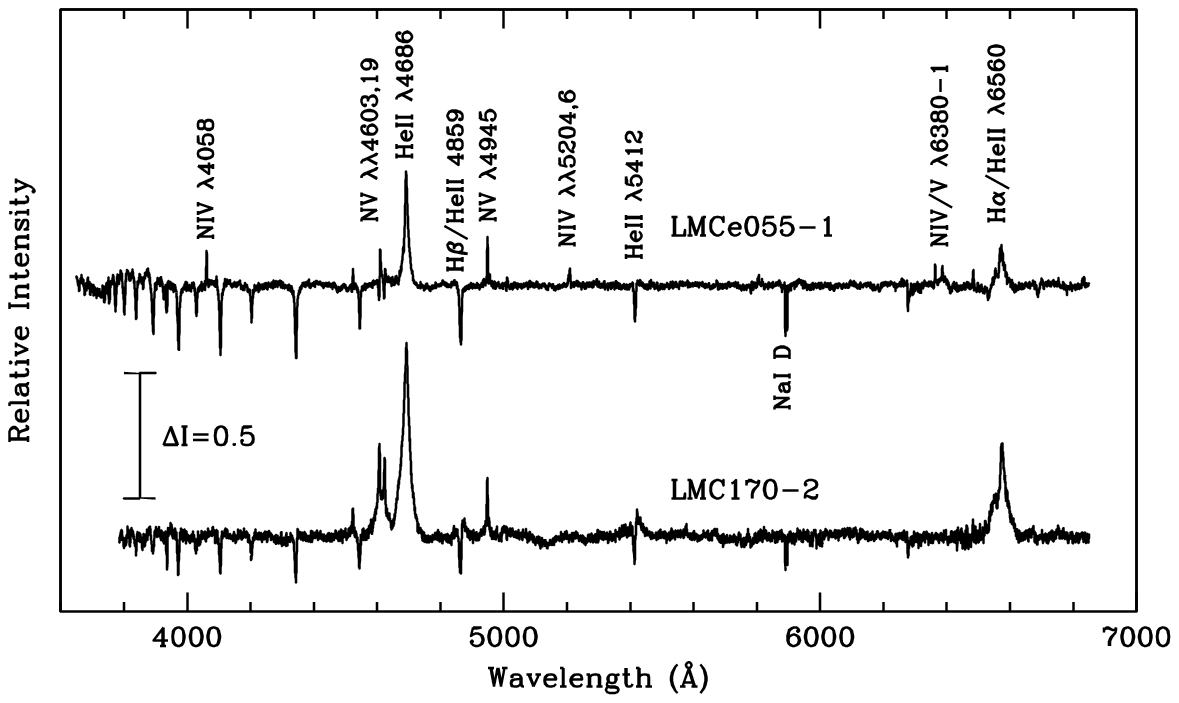}
\vskip -170pt
\plotone{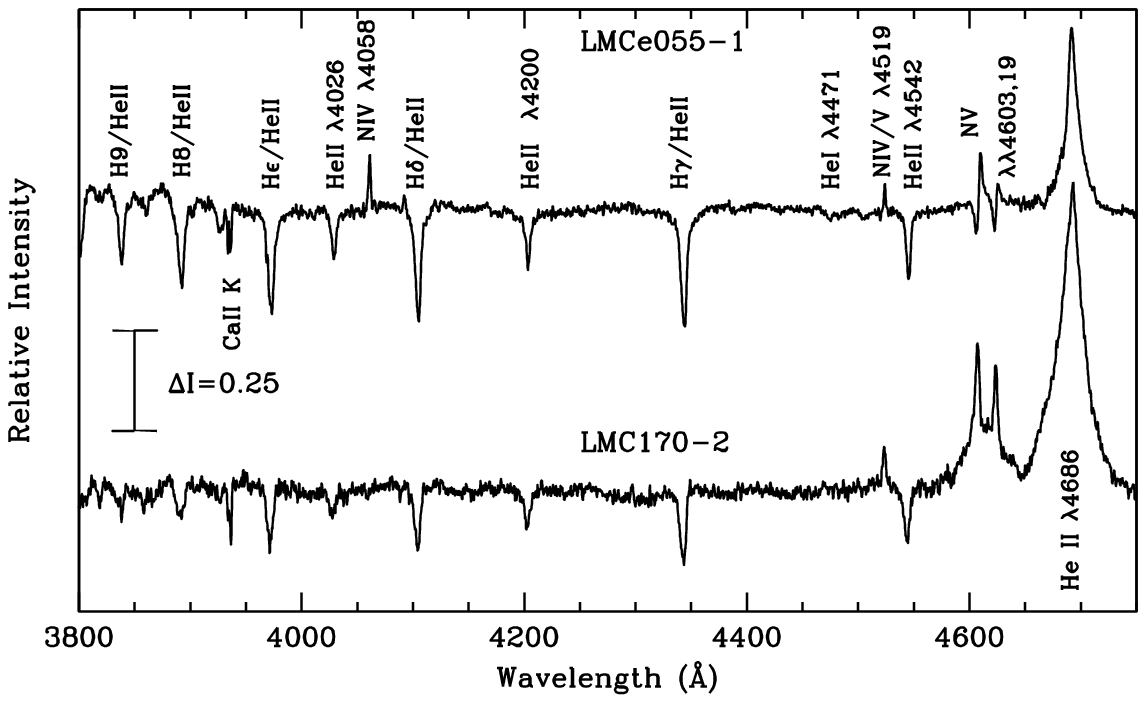}
\caption{\label{fig:lmce0551comp} Comparison between LMCe055-1 and the WN3/O3 star LMC170-2. Note the presence of N\,{\sc iv} emission, the distinctive P Cygni profiles of N\, {\sc v} $\lambda \lambda$4603,19, and the presence of weak He\,{\sc i} $\lambda$4471 absorption in LMCe055-1.  Also note that the even-N Pickering lines of He II, coincident with the Balmer hydrogen lines, are stronger with respect to the odd-N Pickering lines in LMCe055-1, implying a smaller He/H ratio than found in the WN3/O3s.}
\end{figure}

The first impression of the optical spectrum of LMCe055-1 is that it is very similar to that of the WN3/O3s: we see both the emission line spectrum of a high-excitation WN star and an absorption spectrum characteristic of an early O-type star.  Figure~\ref{fig:lmce0551comp} shows a comparison between LMCe055-1 and the WN3/O3 star LMC170-2.  The main differences are that the excitation class is a bit lower than that of a WN3/O3, with the classification lines N\,{\sc iv} $\lambda$4058 similar in height to that of N\,{\sc v} $\lambda \lambda$4603,19, and that there is weak He\,{\sc i} $\lambda$4471 present.  This led \citet{MasseyMCWRIII} to classify it as a WN4/O4.  At the same time, the star cannot be a WN4+O4~V binary: like the WN3/O3s, the star is much too faint to harbor an early O star. As shown in  Figure~\ref{fig:phot}, $V$ is $\sim$16.17.  We will shortly show from our modeling that the reddening of this object is typical of other early-type stars in the LMC with $E(B-V)=0.125$, leading to an absolute visual magnitude $M_V=-2.75$.   Although this is typical for WN3/O3 stars (see Table 2 in \citealt{NeugentWN3O3s}), it is slightly fainter than what is expected for ``normal" WNs:  \citet{2020MNRAS.493.1512R} find that Galactic WN3-4 stars have a continuum V-like absolute magnitude of $-3.6\pm0.5$.  It is many times fainter than what is typical for an O4~V star,  $M_V\sim -5.5$ (see Tables 1-19 and 1-20 in \citealt{1988NASSP.497.....C}), precluding any possibility this is a WN4+O4~V binary.  Thus our initial impression was that LMCe055-1 was a WN3/O3 but of slightly lower temperature, as suggested both by the presence of N\,{\sc iv} emission and He\,{\sc i} absorption, with the companion apparently invisible spectroscopically despite the fact that the secondary eclipse was comparable to the primary eclipse.

As our modeling and radial velocity studies began, it became apparent that this initial impression was flawed.  First, we were easily able to modify our WN3/O3s {\sc cmfgen} models to reproduce most of the spectral features.  The glaring exception was the He\,{\sc i} $\lambda$4471 line.   The N\,{\sc v} and N\,{\sc iv} emission features required an effective temperature\footnote{Technically, the temperature at an optical depth $\tau=2/3$.} of 65,000~K.  But to produce even weak He\,{\sc i} $\lambda$4471 absorption with a similar model required a temperature of 55,000, and there is virtually no N\,{\sc v} emission at that temperature.   This immediately suggested that the WR-like object was actually a ``WN4/O3" and that the He\,{\sc i} $\lambda$4471 absorption came from a companion star of lower temperature. 
Second, even though the WR component (by which we mean the WR emission features plus the H and He\,{\sc ii} absorption) showed
some slight radial velocity variations over the course of time, the radial velocities on consecutive nights were essentially constant, despite being taken at phases corresponding to the eclipse quadrature (i.e., phases of 0.25 and 0.75), when the orbital motion changes should show the greatest differences.  By contrast, the He\,{\sc i} $\lambda$4471 absorption changed by nearly 200 km~s$^{-1}$ on the same consecutive nights.  Naively we thought this merely suggested that there was a large mass ratio between the two components, with the WR star showing little motion because it was so much more massive than its companion.  

As we analyzed our measurements we ran into an insurmountable problem with this interpretation as well: the radial velocity of the He\,{\sc i} $\lambda$4471 line was at its most negative (coming towards us) at a phase of 0.25, and most positive (going away from us) at a phase of 0.75.  However,  the primary (deeper) eclipse must occur when the hotter star is being occulted.   If that is defined as phase 0, then the cooler star will be receding from us (most positive radial velocity) at a phase of 0.25 and coming towards (most negative radial velocity) at a phase of 0.75.  Thus, the eclipse curve could not correspond to the WR star as the primary and the He\,{\sc i} $\lambda$4471 component as the secondary.   This presented quite a conundrum, forcing us to consider an alternative: that we were actually looking at a triple system. 

This epiphany solved several problems.  First, we had been worried about the appearance of the light-curve.  The shape of the light-curve (Figure~\ref{fig:phot}) is that of a nicely detached system with the components well within their Roche lobes.  This seemed highly inconsistent with a massive WR star being one component in a short 2.2-day orbit.  Furthermore, it would explain why the radial velocity of the WR component was essentially constant night-to-night but showed variations of several tens of km~s$^{-1}$ over time, and why the radial velocity of the He\,{\sc i} $\lambda$4471 also showed longer term variations. We expand upon these problems in Section~\ref{Sec-system}. 

In this section we will describe our modeling of the spectrum, as well as the measurements of the radial velocities.

\subsection{Modeling}
\label{Section-modeling}

The goal of our modeling was to determine the physical properties of the WR component. 
To model the UV and optical spectra of LMCe055-1, we used {\sc cmfgen}, a code which solves the radiative transfer and statistical equilibrium equations in spherical geometry, producing a synthetic spectrum \citep{1998ApJ...496..407H}.  Many improvements have been made to the code over the years (\citealt{2012IAUS..282..229H}; see also the appendix in \citealt{ErinWC}). The code is particularly well suited to hot massive stars with stellar winds, but can also be used to analyze other objects of astrophysical interest, such as supernovae (see, e.g., \citealt{10.1111/j.1365-2966.2012.21192.x, 2015MNRAS.449.4304D}).  \citet{NeugentWN3O3} had used the {\sc cmfgen} to analyze the complete sample of WN3/O3 stars, and we began with one of those models in order to analyze the spectra of LMCe055-1.  

\citet{NeugentWN3O3} had found that the physical properties of WN3/O3s differed from that of other WNs primarily in that their mass-loss rates were more similar to that of O-type stars than WNs, with $\log \dot{M} \sim -6$, where the mass-loss rate $\dot{M}$ is in solar masses per year. The WN3/O3s were also on the hot side of what is normal for high-excitation WNs, with effective temperatures of 100,000~K.  Finally, they were unusual in having He/H ratios of $\sim$1.  Although there are certainly WNs with similar He/H ratios, most high-excitation WNs have He/H ratios of 10 or greater. Their CNO abundances were normal, with N mass fractions of about 0.006-0.011, consistent with the surface having a composition reflective of nuclear CNO equilibrium.

The WN3/O3 modeling complexities described by \citet{NeugentWN3O3}  were applicable here: the parameters interact, as both photospheric (absorption) and wind (emission) lines need to be fitted simultaneously.  The derived fit is thus sensitive not only to the usual parameters in modeling WN stars, but also the surface gravity.  We ran {\sc cmfgen} models, varying the luminosity $\log L/L_\odot$, the effective
temperature $T_{\rm eff}$,  the mass-loss rate ($\dot{M}$), the surface gravity ($\log g$), the terminal velocity of the wind ($v_\infty$), the He/H number ratio, and the mass fractions of N, C, and O.  The hydrostatic structure below the sonic point was updated three times during the first 30 iterations; most models were allowed to run for 100 or more iterations.  We adopted a value of 1 for the $\beta$ wind acceleration parameter \citep{1979ApJS...39..481C,1997A&A...323..488S} consistent with \citet{NeugentWN3O3} and studies of other WN stars (see, e. g., \citealt{PotsLMC} and the theoretical arguments put forth in  \citealt{2007ASPC..367..131G}). Once we had our preliminary model, we did explore the effects of varying $\beta$ from 0.8 to 2.0, as discussed briefly below.  Similarly, we set the clumping filling factor to 0.1.  For the non-CNO metals, we set the abundances of Ne, Mg, Si, P, S,  Cl, Ar, Ca, Fe, and Ni to half solar (based on scaling the H\,{\sc ii} oxygen abundances to solar; see discussion in \citealt{RSGWRs}) and ignored the rest.  

In our fitting we used normalized versions of the optical data, and left the UV data in flux units.  We settled on a projected rotational velocity of 110 km~s$^{-1}$  based primarily on a comparison of our models with N\,{\sc v} $\lambda$4945, one of the more narrow features (absorption or emission) in our spectrum.  The models were convolved to the instrumental resolutions of the data before comparison, i.e., a resolving power of 4100 for the optical, and to a resolution of 0.55~\AA\ for the G140L UV data, and 0.09~\AA\ for the two ULLYSES (G130M and G160M) spectra. 

Despite the complexities, some parameters were easily set.  The terminal velocity of 2300 km~s$^{-1}$ is well determined by the blue edge of the P Cyg absorption components of the UV resonance lines, e.g., N\,{\sc v} $\lambda$1240, and C\,{\sc iv} $\lambda \lambda$1548,52.  (The ULLYSES higher resolution UV spectra were particularly useful in establishing this value for C\,{\sc iv}.)  
The optical N\,{\sc iv} and N\,{\sc v} lines helped to initially set the temperature to 60,000-70,000~K; we then adjusted the luminosity to obtain good agreement with the UV flux after correcting for interstellar reddening. Following our previous work (e.g., \citealt{NeugentWN3O3,ErinWC,ErinWO}) we assumed a foreground reddening of $E(B-V)$=0.08 (based on the dust maps of \citealt{Sch}) using a \citet{CCM} law, and added in LMC reddening correction of $\sim$ 0.05~mag using a \citet{1983MNRAS.203..301H} law. Together the temperature and luminosity then fix the radius. The mass-loss rate interacts with luminosity, temperature, surface gravity, and abundances (sometimes in unexpected ways) to produce the line fluxes.  

An obvious complication was deciding how to allow for the contribution of the secondary (and other) components.    Given that the secondary star was cool enough to produce He\,{\sc i}, it seemed likely that its contribution in the UV would be negligible, and its affect in the optical could then be evaluated from the model.  

\begin{figure}
\epsscale{1.0}
\plotone{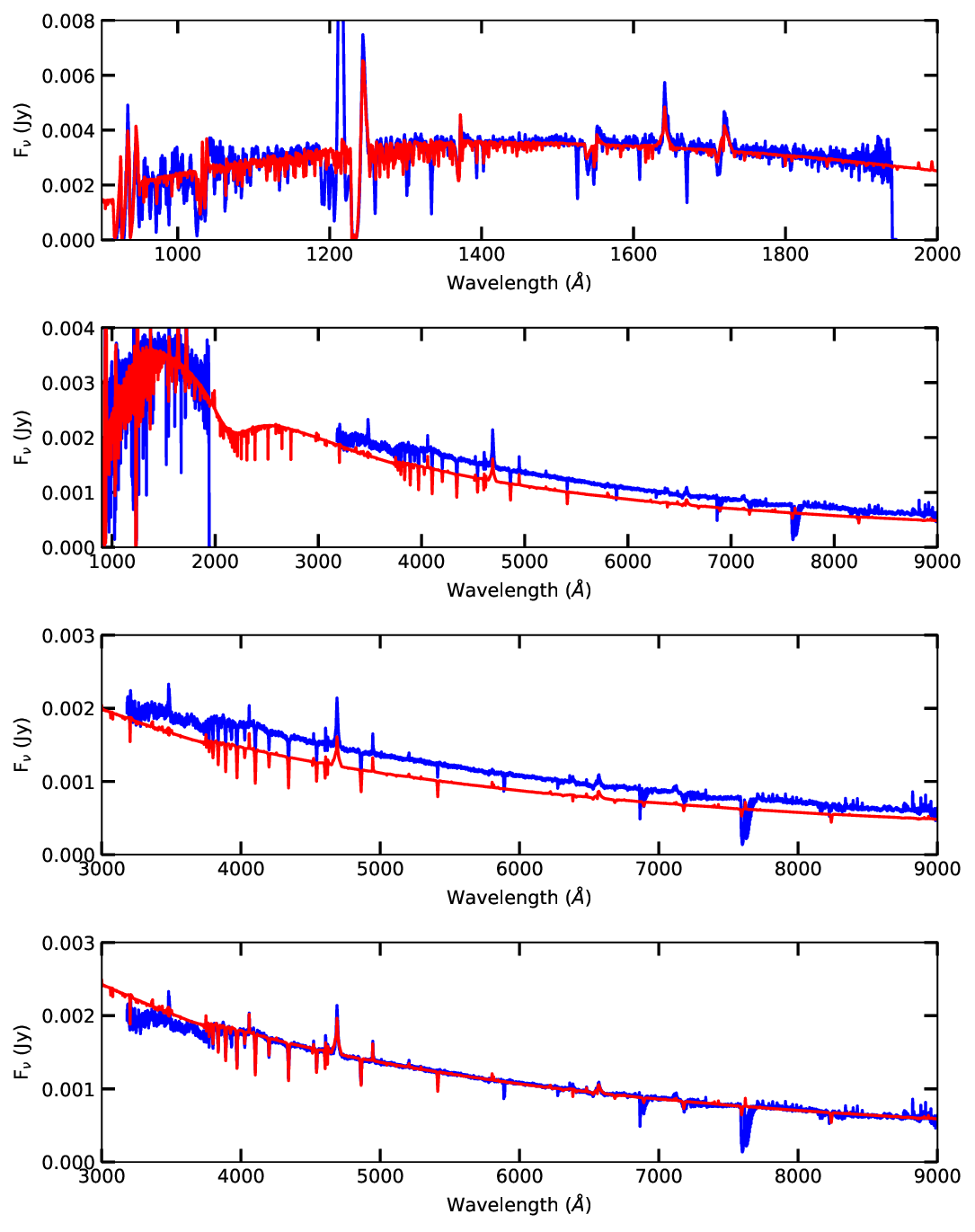}
\caption{\label{fig:SED} Spectral energy distribution (SED).  The SED of the model is shown in red; the data are shown in blue. The model flux has been reddened using a combination of a Galactic \citet{CCM} reddening law with $E(B-V)=0.08$ and a  \citet{1983MNRAS.203..301H} LMC reddening law with $E(B-V)=0.055$.  The data have been corrected to vacuum wavelengths and to the rest frame assuming a 250 km~s$^{-1}$ radial velocity determined by fitting our optical spectral lines. (This radial velocity value is consistent with the LMC's systemic radial velocity of 262 km s$^{-1}$ [\citealt{2002AJ....124.2639V}]; see also \citealt{NeugentLMC}.)  The top panel shows the excellent match with the UV spectrum. The middle two panels show that the optical spectra shows additional flux contribution from the companion star(s). In the bottom panel the model has been scaled by a factor of 1.22 to produce good agreement with the optical.  Note that the optical flux below 3600~\AA\ suffers from calibration issues. }
\end{figure}

In Figure~\ref{fig:SED}  we show a comparison between the model SED (red) and the observed (blue) fluxes.  We have aimed the model to agree in flux with the UV data, and in the top panel we see we have succeeded.  In the middle panels we see that the optical data show more flux than we would expect.  In the bottom panel we see we can achieve excellent agreement with the optical SED by scaling the model flux by a factor of 1.22.  Thus, our conclusion is that the companion(s) stars contaminate the optical data by about 18\%, but that 82\% of the combined optical flux is due to the WR component. 

With this new information we can estimate the absolute magnitudes of the components.  The total system has an absolute  visual magnitude $M_V=-2.75$.  Given that we had to multiply the flux of the model by 1.22, this implies an $M_V$(WR)=$-2.5$, which is also the average absolute magnitudes of the WN3/O3 stars \citep{NeugentWN3O3s}.   The remaining component(s) then have an $M_V=-0.9$.  Given that we see (very weak) He\,{\sc i} $\lambda$4471, the brighter of the secondary stars must be earlier than A0.  The absolute magnitude of a B5~V is -1.2 and that of a B8~V is -0.25 (from Table 15.7 of \citealt{Allen}).   These values are consistent with the conclusions of  Section~\ref{Sec-system}, where we argue that the He\,{\sc i} $\lambda$4471 component is approximately of  B7~V type; i.e., our proposed model of the system is at least self-consistent.

We have therefore computed {\sc cmfgen} models with effective temperatures of 12,000-15,000 and $\log g = 4.0$, corresponding to a mid-to-late B type dwarf using LMC-like metallicities (i.e., half solar).  (A comparison with the TLUSTY BSTAR2006 models of \citealt{2007ApJS..169...83L}  showed only negligible differences in the normalized spectra.)  If the star is tidally locked with a 2.2 day period, then the expected rotational velocity is $\sim$80~km~s$^{-1}$, and we broadened the model spectra to this value as well as matching the instrumental resolving power $R\sim 4100$. We then multiplied the normalized, broadened spectrum by a factor of 0.18, subtracted it from our normalized optical MagE spectrum, and renormalized it by multiplying by 1.22. (The wavelength scale was first adjusted from vacuum to air, and shifted to 250 km~s$^{-1}$.)  Obviously this correction procedure is approximate, but it provides a more realistic spectrum to model.  There was little differences between the results of these models, and we (somewhat arbitrarily) adopted the 15,000 K model.

The result of this correction was to weaken the hydrogen Balmer lines of the WR component considerably.  Visually the (combined) spectrum in Figure~\ref{fig:lmce0551comp} looked as if the hydrogen to helium ratio was almost normal, and indeed our original {\sc cmfgen} modeling indicated a He/H ratio of 0.15 rather than the $\sim$1 found by \citet{NeugentWN3O3} for the WN3/O3s.  Even though the B star contribution is relatively minor, the hydrogen lines are strongest in late B/early A stars, and thus removing the B star results in a He/H number ratio closer to what we expect.
We quickly found that a He/H ratio of 2 was a much better fit than 0.15!  The effect on the other lines were primarily to make them stronger by $\sim$20\% as extraneous continuum was removed, and indeed the changes in the other physical parameters were minor.   Before
the correction, we could not obtain an adequate fit to the wings of H$\gamma$ with realistic surface gravities. Removing the B star, we found a $\log g=4.4-4.5$.   The effective temperature was unchanged as it is largely fixed by the  N\,{\sc v} to N\,{\sc iv} ratio, which was not affected by the correction. The SED of the new final model is indistinguishable from the preliminary fit shown in Figure~\ref{fig:SED}, and we did not iterate the process further.


\begin{deluxetable}{l c c}[h]
\tablecaption{\label{tab:modeling}Physical Properties of the WR Component}
\tablewidth{0pt}
\tablehead{
\colhead{Property}
&\colhead{LMCe055-1}
&\colhead{LMC170-2}
}
\startdata
$T_{\rm eff}$ [kK]        &65             & 100    \\
$\log L/L_\odot$       & 4.97          & 5.69 \\
$R_*/R_\odot$           & 2.35           & 2.2    \\
$R_{2/3}/R_\odot$.    & 2.40           &2.3   \\
$\log g$ [cgs]            &4.5            & 4.95  \\
$\log \dot{M}$            &-6.52        & -5.82 \\
$v_\infty$ [km~s$^{-1}$] & 2300      & 2600 \\
He/H [by number].     & 2.0            & 1.0   \\
C abund mass fract  & 1E-4    & 1E-4 \\
N abund mass fract  & 0.011       & 0.011 \\
O abund mass fract  &  2E-5    & 8E-5 \\
Mass ($M_\odot$)     & 7            & 15 \\
$M_V$                       &-2.5        & -2.9 \\
\enddata
\tablecomments{For both the LMCe055-1 modeling here, and the LMC170-2 modeling by \citet{NeugentWN3O3}, a value of $\beta=1$ was
assumed, along with a clumping filling factor of 0.1.}
\end{deluxetable}
In Table~\ref{tab:modeling} we compare the derived physical properties of LMCe055-1 with that found by \citet{NeugentWN3O3} for the WN3/O3 star LMC170-2.  As a reminder, $R_*$ refers to the ``core" radius, the inner boundary of the model atmosphere, where the outflow velocity is negligible (see, e.g., \citealt{1989A&A...210..236S}).  $R_{2/3}$ refers to the radius at which the Rosseland optical depth $\tau$ is 2/3rds; the difference between $R_*$ and $R_{2/3}$ characterizes how extended the atmosphere is. (For BAT99-3, a WN4 star we are analyzing, $R_{2/3}$ is about twice as large as $R_*$; i.e., neither LMCe055-1 nor the WN3/O3s are extended compared to this more normal WN star.)  As a result, the effective temperature and surface gravities at $\tau=2/3$ are very similar to those at the bottom of the atmosphere, $\tau=100$, where the radius is essentially $R_*$.

\begin{figure} [h]
\epsscale{1.0}
\plotone{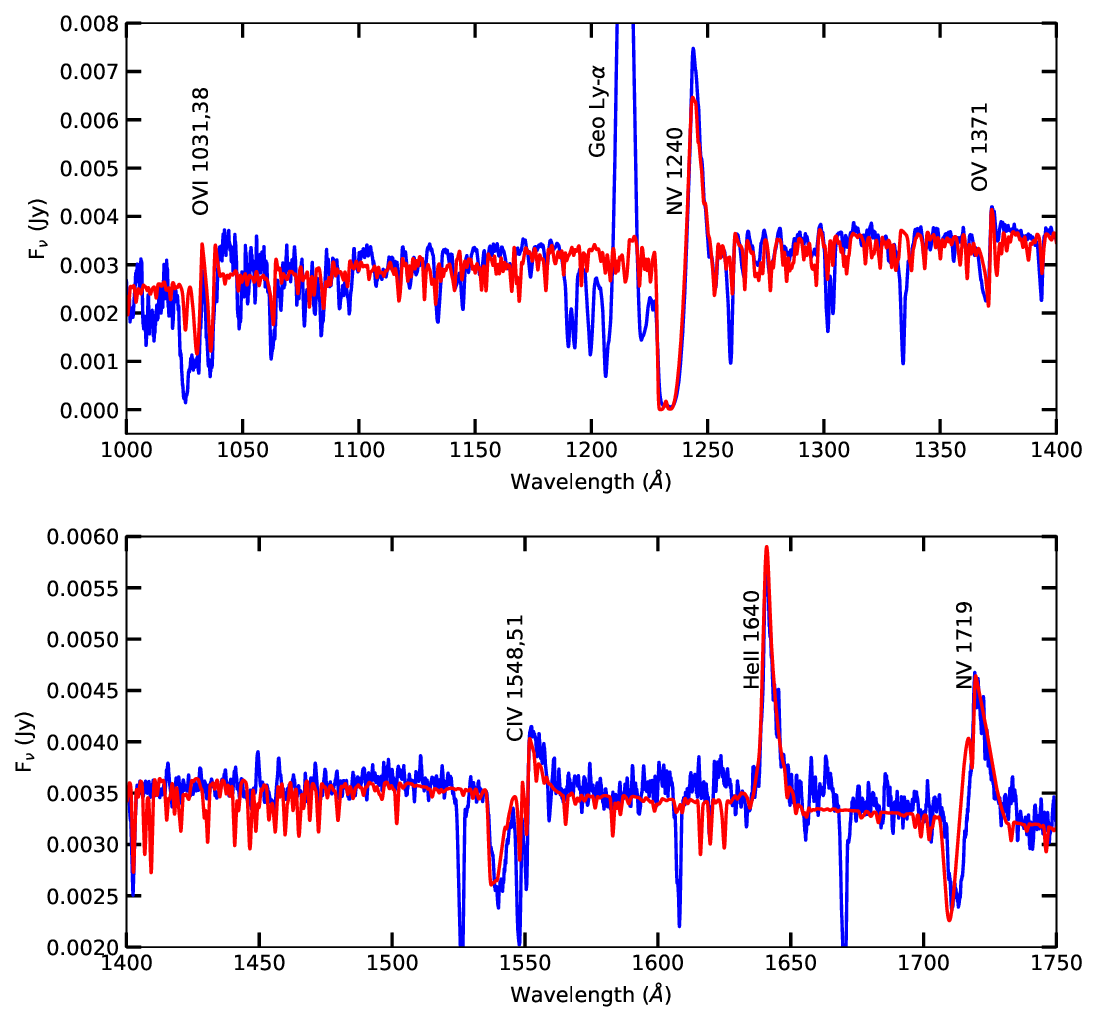}
\caption{\label{fig:UV} Model Fit in the UV.  The best-fitting {\sc cmfgen} model is shown in red, and the G140L/800 COS observation in blue.  The lines used in the fitting are labeled. The model spectrum has been convolved to a spectral resolution of 0.55~\AA\ in order to match the data, and further broadened by a rotation component of 100 km~s$^{-1}$.}
\end{figure}

\begin{figure}
\epsscale{1.2}
\plotone{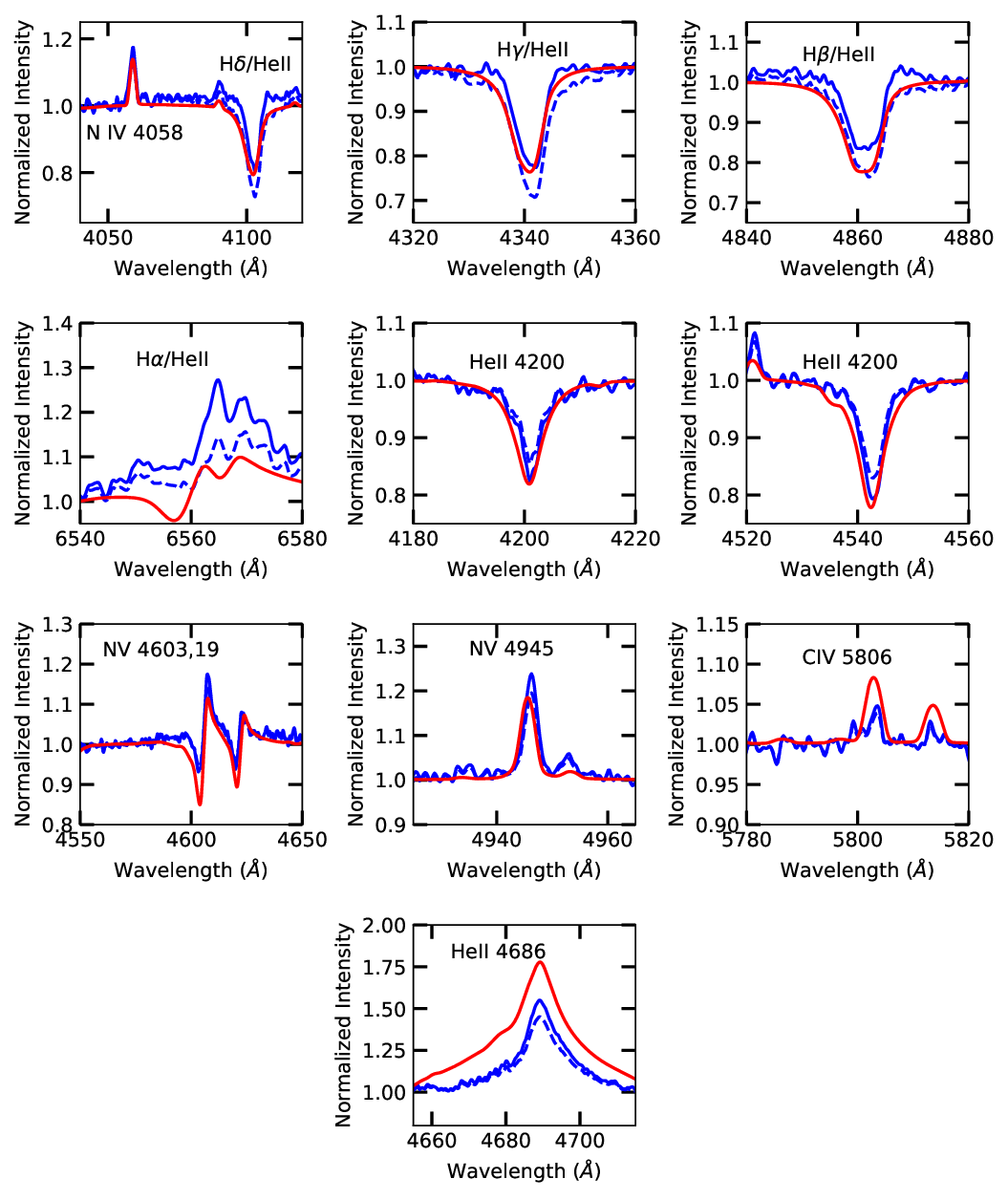}
\caption{\label{fig:opt} Model Fit in the optical for selected spectral lines.  The normalized best-fitting {\sc cmfgen} model is shown in red, and the normalized MagE spectrum  is shown in blue.  The solid blue line represents the spectrum after the contribution of the B star has been removed; the dashed blue line shows the MagE data before this correction.  The model spectrum has been convolved to a resolving power of 4100 and broadened by a rotation component of 100 km~s$^{-1}$.}
\end{figure}

In Figures~\ref{fig:UV} and \ref{fig:opt} we show the match between our best model and the spectral features.\footnote{We ran 40 models in order to obtain the pre-corrected fits, and an additional 20 to refine the parameters once we removed the B star component from the optical spectrum.}  The ``best" model is far from perfect: the observed H$\alpha$ line is much stronger than the model fit, while the He\,{\sc ii} $\lambda$4686 is much weaker.  We could get a much better H$\alpha$ fit by lowering the He/H ratio but this resulted in an unacceptable fit for the hydrogen absorption lines.  Alternatively, we found that increasing $\beta$ from 1.0 to 1.5 resulted in a much better fit to H$\alpha$ but a worse fit to N\,{\sc v} $\lambda\lambda$4603,19, He\,{\sc ii} $\lambda$1640, and
He\,{\sc ii} $\lambda$4686.  The mass loss rate $\dot{M}$=3.0E-7  $M_\odot$ yr$^{-1}$ ($\log \dot{M}=-6.2$) is a relatively good answer for most lines; increasing it to 3.5E-7 gives a much better fit to the C\,{\sc iv} $\lambda\lambda$1548,52 resonance line, while a lower value of 2.5E-7 does a better job of matching He\,{\sc ii} $\lambda$1640.   The oxygen abundance is based purely on obtaining a good fit to the O\,{\sc v} $\lambda$1371 line, as the only other oxygen line present, O\,{\sc vi} $\lambda \lambda$1031, 38, is relatively insensitive to the adopted abundance.   The C\,{\sc iv} $\lambda$5806 line is poorly fit.  Reducing the carbon abundance by a factor of 2 results in a much better fit, but destroys the good agreement with C\,{\sc iv} $\lambda\lambda$1548,52. 
These compromises are more extreme than what was required by \citet{NeugentWN3O3} but here we deal with the uncertain correction for the companion(s).  Uncertainties in the best value of $\beta$ and the filling factor further cloud the issue. 

We do emphasize that although we adopt a He/H number ratio of 2, a value of 1 or even 0.5 cannot be ruled out. Although the value of 2 gave a superior fit to the hydrogen/helium absorption lines, their strengths are very sensitive to the correction we made for
the companion star.  

\clearpage

\subsection{Radial Velocities}

The key to understanding the nature of the LMCe055-1 system came from the radial velocities.  Our modeling adequately duplicates all of the observed spectral features with the exception of the weak  He\,{\sc i} $\lambda$4471 line. Although we realized early on in our analysis  that this line had to be due to a second star, it was the analysis of the radial velocities that convinced us that we were dealing with a higher-order multiple system, and that the WR component played no role in the 2.2 day period shown by the eclipses.  

Cooler stars offer a rich plethora of narrow spectral features allowing radial velocity measurements on the order of 1~km~s$^{-1}$ with data such as ours.  In contrast, O-type stars have only a handful of spectral lines, with rotational broadening of 100 km~s$^{-1}$ at least.   WR emission lines are broadened to $\sim$1000 km$^{-1}$ or more, due to their formation in outflowing winds.  Recently \citet{2023ApJ...947...77M} achieved good results in placing constraints on the binarity of the WN3/O3s by using cross-correlation of carefully selected lines.  In investigating the best method to use here we were aided by the high S/N spectra obtained on 2018 February 4 and 2018 February 5.  On each night we obtained two consecutive 1-hr exposures, each with its own independent comparison spectrum (see Table~\ref{tab:spectra}).  Assuming no radial velocity changes over a 2-hr period, we could then see what method resulted in the best agreement between the two consecutive exposures.  We adopted a method similar to that of \citet{2023ApJ...947...77M}, except that rather than cross-correlating each spectrum against each other, we used the model spectrum as our template. (The model used was an intermediate version, the output of which was only subtly different than our final adopted model.)  The model template, of course, had essentially infinite S/N and was much finer resolution.  To prepare for the cross-correlation, we normalized both the model and the observed spectra to the continuum and subtracted unity so that the continuum made no contribution.  The model spectrum was degraded to the same resolution as our observed spectra, and converted from vacuum to air wavelengths. 

The lines which gave the most consistent velocities were the strongest absorption features,
H$\delta$, H$\gamma$, H$\beta$, and He\,{\sc ii} $\lambda$4542, and the emission features
N\,{\sc iv}$\lambda$4058 and the N\,{\sc v} $\lambda \lambda$4603,19 / He\,{\sc ii} $\lambda$4686 complex. The cross-correlation was computed using the ``fxcor" function within {\sc iraf}, fitting the cross-correlation peak with a Gaussian.  In averaging the
results we compared the mean, the median, and the ``edited mean", obtained by discarding the highest and lowest
value before computing the mean.  We believe the latter gives the most robust measure.

As for the broad, weak He\,{\sc i} $\lambda$4471 feature, we measured the radial velocity ``by hand," marking the
continuum by eye and computing both the centroid and fitting the line by a Gaussian function with {\sc iraf}.  We only measured the line on the best spectra, and even so the measurements were very uncertain.  This is evident by the large differences between the two methods.  We an assign an uncertainty of 30 km s$^{-1}$ to the He\,{\sc i} velocities.

We give the heliocentric radial velocity measurements in Table~\ref{tab:RVs}.   We repeat the eclipse phases from Table~\ref{tab:spectra},  computed from our revised ephemeris in Section~\ref{Sec-epherm}.
\begin{deluxetable}{l c c c c c c c c c}
\tabletypesize{\scriptsize}
\tablecaption{\label{tab:RVs}Radial Velocities}
\tablewidth{0pt}
\tablehead{
\colhead{HJD-2450000}
&\colhead{Phase}
&\multicolumn{5}{c}{WN4 Component (km~s$^{-1}$)}
&
&\multicolumn{2}{c}{$\lambda$4471 Component (km~s$^{-1}$)} \\
\cline{3-7} \cline{9-10} 
&&\colhead{Mean}
&\colhead{$\sigma$}
&\colhead{Median}
&\colhead{Edited Mean}
&\colhead{$\sigma$}
&&\colhead{Gaussian}
&\colhead{Centroid} 
}
\startdata
7398.607   &   0.117   &    244.3   &   3.5   &   244.6   &   244.1   &   3.1   &&   \\
7792.573   &   0.589   &    267.0   &   2.6   &   264.7   &   265.2   &   1.2   &&   228.0   &   257.7   \\
7792.698   &   0.647   &    263.4   &   2.1   &   262.0   &   262.7   &   1.6   &&   303.0   &   352.9   \\
7793.691   &   0.107   &    260.8   &   5.0   &   262.4   &   261.3   &   1.5   &&   190.5   &   190.7   \\
8118.552   &   0.572   &    251.7   &   5.3   &   250.3   &   249.3   &   1.5   &&   336.7   &   374.2   \\
8119.553   &   0.036   &    242.2   &   3.7   &   238.9   &   240.6   &   3.1   &&   256.3   &   284.9   \\
8124.552   &   0.351   &    239.5   &   5.6   &   235.2   &   238.4   &   4.7   &&   \\
8153.541   &   0.778   &    250.5   &   4.9   &   247.0   &   249.5   &   3.6   &&   \\
8153.574   &   0.793   &    244.5   &   5.0   &   242.7   &   242.8   &   3.5   &&   \\
8153.556\tablenotemark{a}   &   0.785   &    248.0   &   4.6   &   246.8   &   247.0   &   1.9   &&   346.5   &   332.7   \\
8154.577   &   0.258   &    225.2   &   4.8   &   226.5   &   224.1   &   4.0   &&   \\
8154.621   &   0.278   &    234.1   &   4.1   &   229.4   &   232.0   &   2.8   &&   \\
8154.598\tablenotemark{a}   &   0.268   &    229.6   &   4.2   &   226.8   &   227.5   &   3.2   &&   \\
8440.641   &   0.754   &    244.2   &   7.9   &   239.5   &   239.1   &   1.6   &&   408.1   &   360.1   \\
8482.748   &   0.256   &    246.1   &   4.2   &   243.0   &   243.2   &   1.9   &&   226.5   &   194.9   \\
8864.789   &   0.205   &    245.5   &   7.6   &   241.4   &   242.8   &   4.0   &&   \\
9187.668   &   0.753   &    265.6   &   3.9   &   266.7   &   266.1   &   1.1   &&   375.4   &   397.9   \\
9253.557   &   0.270   &    250.5   &   6.0   &   243.4   &   248.3   &   5.5   &&   136.8   &   154.3\\
9254.585   &   0.746   &    248.5   &   3.6   &   249.5   &   247.8   &   2.1   &&   320.6   &   347.4   \\
9854.829   &   0.760   &    247.1   &   4.2   &   249.9   &   247.3   &   4.1   &&   488.9   &   474.8   \\
\enddata
\tablenotetext{a}{Combined from previous two spectra.}
\end{deluxetable}

\section{The LMCe055-1 ``System"}
\label{Sec-system}

\subsection{Overview}

In previous sections we have described our data and our analysis.  Here we now put these pieces together to describe the LMCe055-1 system.

\begin{figure}
\epsscale{0.6}
\plotone{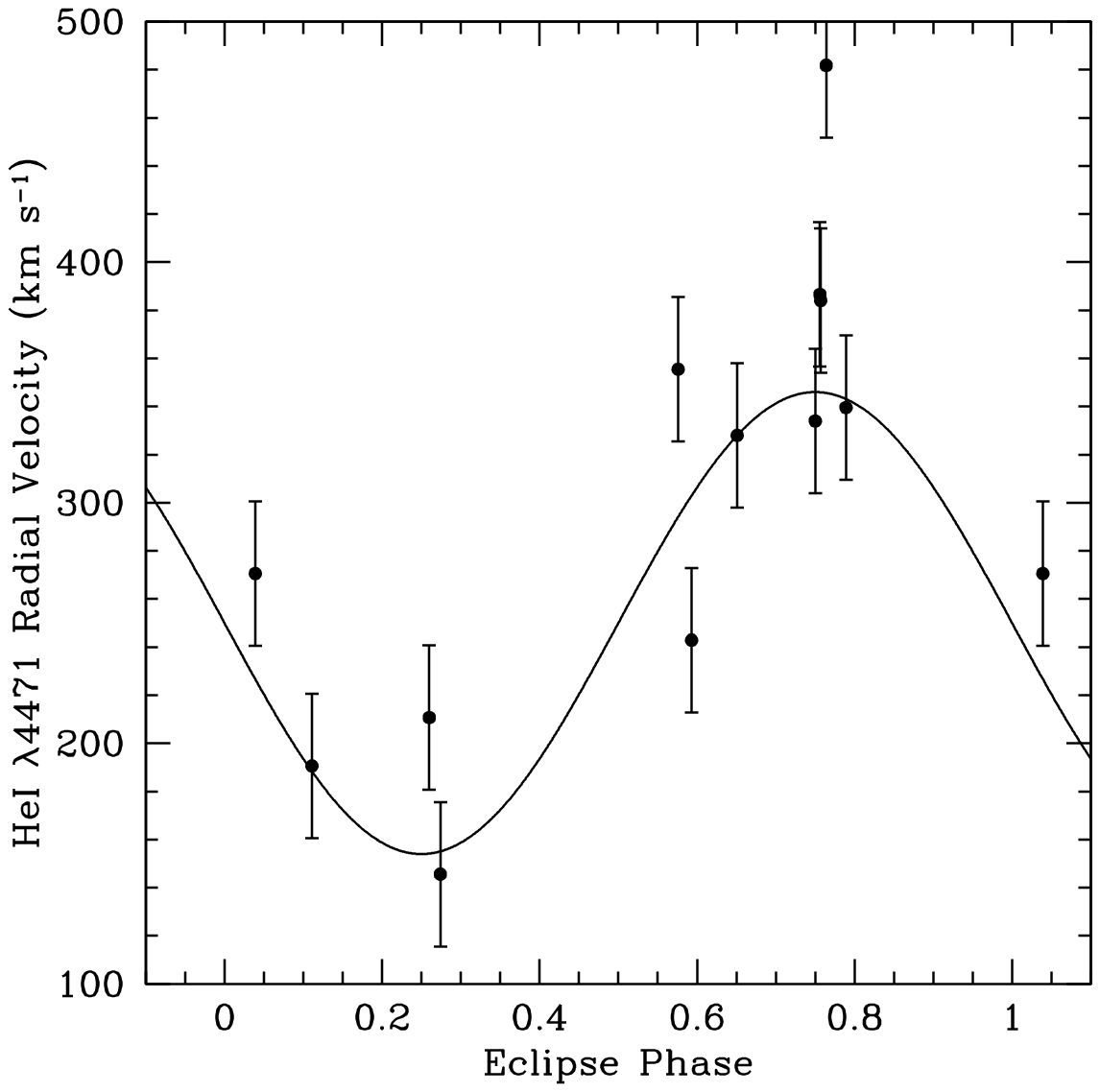}
\caption{\label{fig:4471RV}  The radial velocities of the He\,{\sc i} $\lambda$4471 component.  The radial velocities of the He\,{\sc i} $\lambda$4471 component (averaged of the two methods given in Table~\ref{tab:RVs}) are shown as a function of the eclipse phase, with phase 0 being the primary eclipse.  If the system consisted of the WR star as the primary, and the He\,{\sc i} $\lambda$4471 component as the secondary, then we would expect the latter to have the largest velocities at a phase of 0.25 as it moved away from us after passing in front of the hotter WR star.  Instead, the opposite is seen, suggesting that He\,{\sc i} $\lambda$4471 arises in the hotter primary of the 2.2-day pair, with an otherwise undetected third component. The error bars have been set to $\pm$30 km~s$^{-1}$, a rough estimate of our measuring uncertainty. The solid black line shows the best-fit radial velocity curve with  $\gamma$ fixed to 250~km~s$^{-1}$, our preferred solution.}
\end{figure}

Our notion that the WR star and the He I $\lambda$4471 component represented the primary and secondary of the 2.2 day eclipsing pair met two irrefutable problems, as previously mentioned.  The WR star was clearly the hotter star, given its 65,000 K 
temperature; a much cooler photosphere (11,000-45,000~K) is needed to produce He\,{\sc i} $\lambda$4471 absorption.  The deeper, primary eclipse must happen when the hotter star is covered.  The secondary should then have its most positive radial velocity (moving away from us) at phase 0.25, and its most negative radial velocity at phase 0.75.  But even a casual inspection of the He\,{\sc i} radial velocities in Table~\ref{tab:RVs} shows that the opposite is true: the He\,{\sc i} velocities are the largest at phase 0.25 and smallest at phase 0.75.  We illustrate this in Figure~\ref{fig:4471RV}. Thus primary eclipse is occurring when the He\,{\sc i} source is being covered.  This requires a third component to be present in a 2.2-day orbit with  the He\,{\sc i} source.

The second problem is that the radial velocities of the WR show almost no changes from night-to-night. Yet, during the course of this study, the radial velocities of the WR component range from 224 to 265 km~s$^{-1}$, far greater than the 1 to 5 km~s$^{-1}$ uncertainty in their measurement. But the time-scale for this motion is clearly inconsistent with a 2.2 day period.  For instance, consider the two Gemini GMOS spectra taken at phase 0.27 and 0.75 on two consecutive nights (2459253.557  and 2459254.585).  Their velocities agree to 0.5 km$^{-1}$!  

With this in mind, we now treat the system as (at least) a triple. We will return to the issue of whether this is or not a {\it physical} multiple below.

\subsection{The 2.2 Day Binary}
\label{Sec-2.2bin}

\begin{figure}
\epsscale{0.5}
\plotone{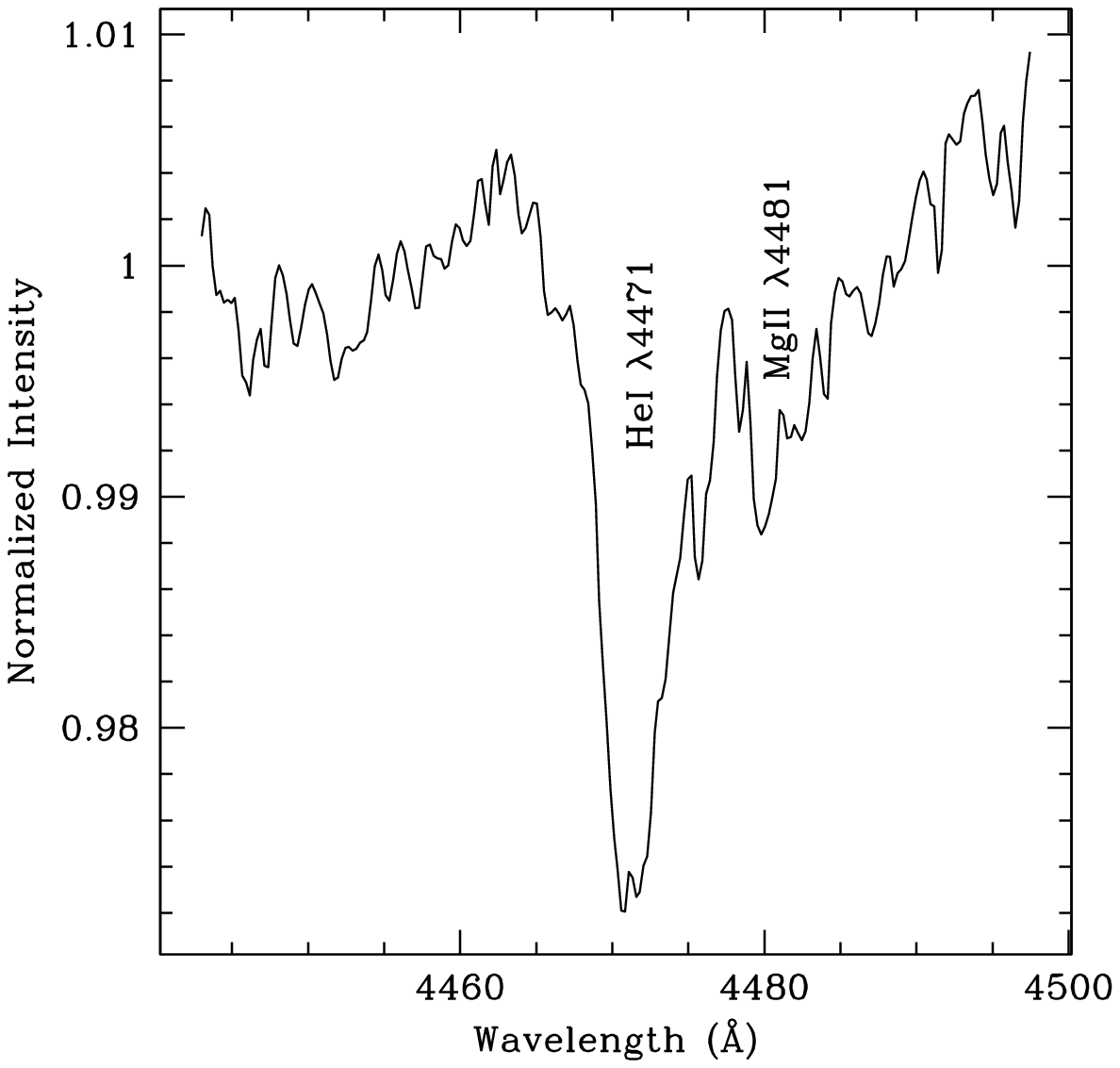}
\caption{\label{fig:4471Spect} Composite He\,{\sc i} $\lambda$4471 line. By shifting the spectra by the radial velocities to a rest frame, and combining them with a sigma clipping algorithm, we recover the weak He\,{\sc i} $\lambda$4471 line.  Even the neighboring Mg\,{\sc ii} $\lambda$4481 is seen.  Its strength, relative to the He\,{\sc i} lines, leads to a B7~V classification.}
\end{figure}

Our analysis of the 2.2-day binary begins with the determination of the spectral type of the He\,{\sc i} component  (henceforth ``the primary of the 2.2-day binary"). We know from the lack of velocity shift of the He\,{\sc ii} absorption lines that the 2.2-day primary does not contribute significantly to those lines.  We therefore infer that the star is later than an O-type; at the same time it can not be as late as an A0~V since He\,{\sc i} is present. Thus the primary of the 2.2-day star is some type of B star. However, we can do better than that.  We have used the radial velocities of the He\,{\sc i} component in Table~\ref{tab:RVs} to shift each of those spectra to the rest frame, and then averaged them, using a sigma-clipping algorithm.  We display the
result in Figure~\ref{fig:4471Spect}.  We not only see the He\,{\sc i} $\lambda$4471 feature but possibly also neighboring Mg\,{\sc ii} $\lambda$4481.  We classify this spectrum as B7~V, similar to that shown of HR~1029 in Plate 13 of \citet{MorganAbt}; see also Figures 4.1 and 4.2 of \citet{2009ssc..book.....G}. We expect such a star to have a mass of 4-5$M_\odot$ (see Table 15-8 of \citealt{Allen}). 

Although we do not see spectroscopic evidence of the secondary of the 2.2-day binary we can still learn about it from the mass function of its primary. 
With a short 2.2 day period the orbit is likely circular; this is consistent with the eclipse curve (Figure~\ref{fig:phot}), which shows that the primary and secondary eclipses are separated by precisely 0.5 phase.   We have therefore calculated the best fitting sine curve as an approximation to its orbit, using the light-curve ephemeris.  We give its parameters in Table~\ref{tab:HeIShope}.   We show two solutions for the orbital semi-amplitude $K_p$, one with the center of mass motion $\gamma$ a free parameter, and one with it fixed to 250 km $^{-1}$, the median radial velocity found from Gaia DR3 for stars in the immediate vicinity.  The semi-amplitude of the primary $K_p$ is 117 km s$^{-1}$ if $\gamma$ is free, and 96 km s$^{-1}$ with $\gamma$ fixed.  We can compare these values to what we obtained on the two consecutive nights of Gemini data,  2459253.557  and 2459254.585, at phases of 0.27 and 0.75, respectively.  The difference between these two measurements is $189\pm50$ km~s$^{-1}$.  If the eccentricity is 0, this gives us a reasonable measure of 2$K_p$, $K_p=94.5$ km~s$^{-1}$.  This is in good accord with the $K_p$=96 km~s$^{-1}$ we estimate from the sine curve with $\gamma$ fixed.   We adopt this as our orbit solution for the 2.2-day pair.

\begin{deluxetable}{l c}
\tabletypesize{\scriptsize}
\tablecaption{\label{tab:HeIShope}Orbital Parameters of the 2.2-day Binary}
\tablewidth{0pt}
\tablehead{
\colhead{Property}
&\colhead{Value}
}
\startdata
Primary Spectral Type & B7~V \\
T$_0$ (HJD-245000) & 7001.0906 \\
P (days)    & 2.159044 \\
$M_V$ (mag) & $-0.9$ \\
$\gamma$ (km~s$^{-1}$) & 266 \\
$K_p$   (km~s$^{-1}$) & 117 \\
$f(m)$ ($M_\odot$) & 0.36 \\
$K_p$ (km~s$^{-1}$), $\gamma$ fixed to 250 km~s$^{-1}$  &  96 \\
$f(m)$ ($\gamma$ fixed)  & 0.20 \\
\enddata
\tablecomments{Typical uncertainties in the orbital semi-amplitude $K$ are 10 km s$^{-1}$ but
depend strongly on how the individual data are weighted. 
Note that the last two entries computed with $\gamma$ fixed are the
preferred values.  }
\end{deluxetable}

The mass function $f(m)$ is defined as $$f(m)=1.036 \times 10^{-7}  K_p^3  P  (1-e^2)^{(3/2)} = \frac{m_s^3 \sin^3 i}{(m_p+m_s)^2},$$
where $K_p$ is in km~s$^{-1}$, the $P$ is  in days, $e$ is the orbital eccentricity, $i$ is the orbital inclination, and $m_p$ and $m_s$ are the masses of the primary and secondary, respectively.\footnote{There is a nice derivation of this in \citet{Smart}, although the constant given, 1.038E-7, differs slightly than what is given in more modern sources, such
as \citet{Allen}.  We were amused to find that the discrepancy is due to the change in our knowledge of the size of an
astronomical unit!    \citet{Smart} assumed that 1 AU was 149,500,000 km,
rather than the current IAU value of 149,597,870.7 km. Since the constant depends upon the size of an AU to the inverse third power, this changes the 1.038E-7 value to 1.036E-7.  We note that the units on this are quite a jumble, with the period expressed in days, and the velocity in km per second.}  We have included the mass functions in Table~\ref{tab:HeIShope}.

\begin{figure}
\epsscale{0.45}
\plotone{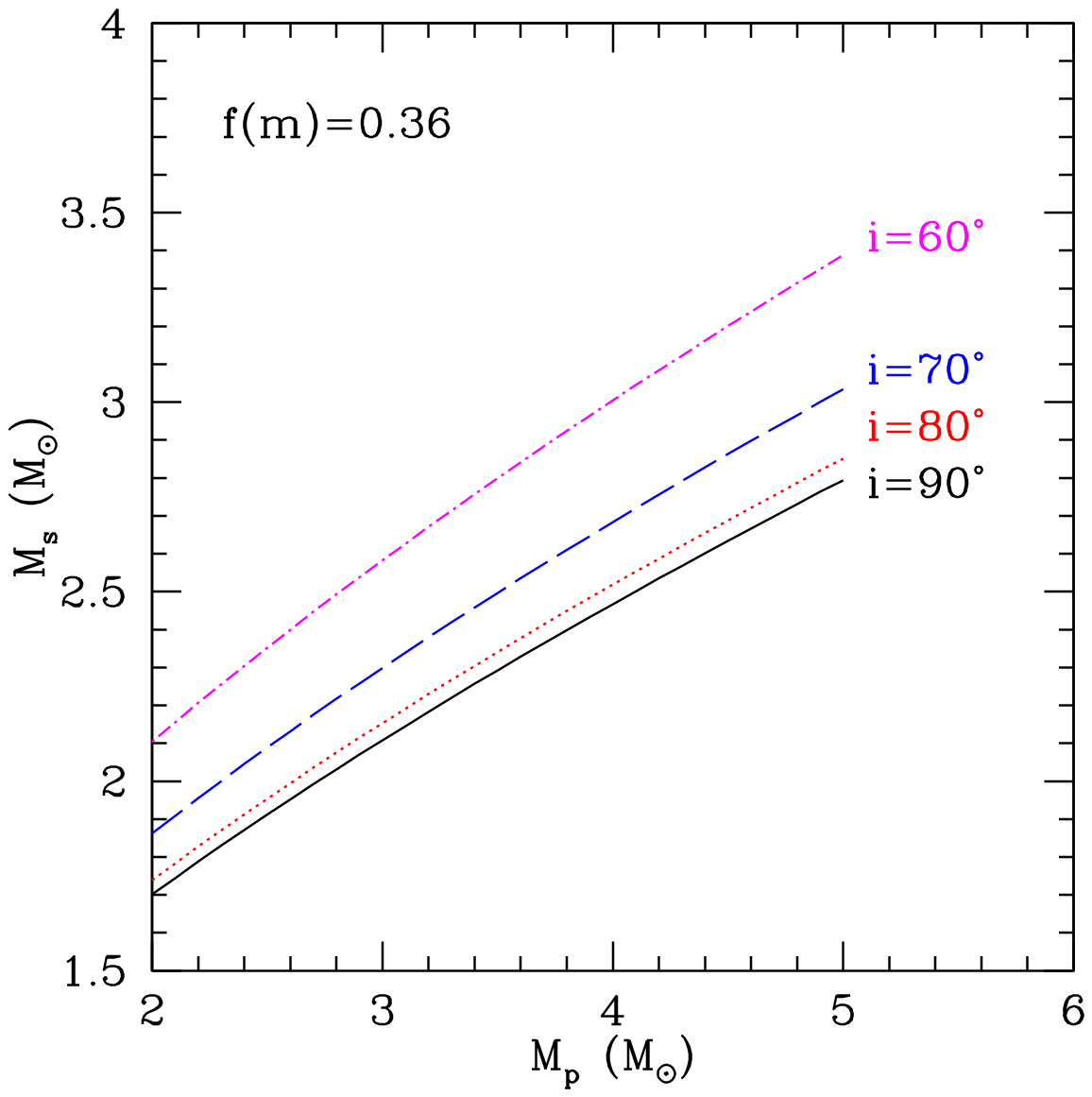}
\plotone{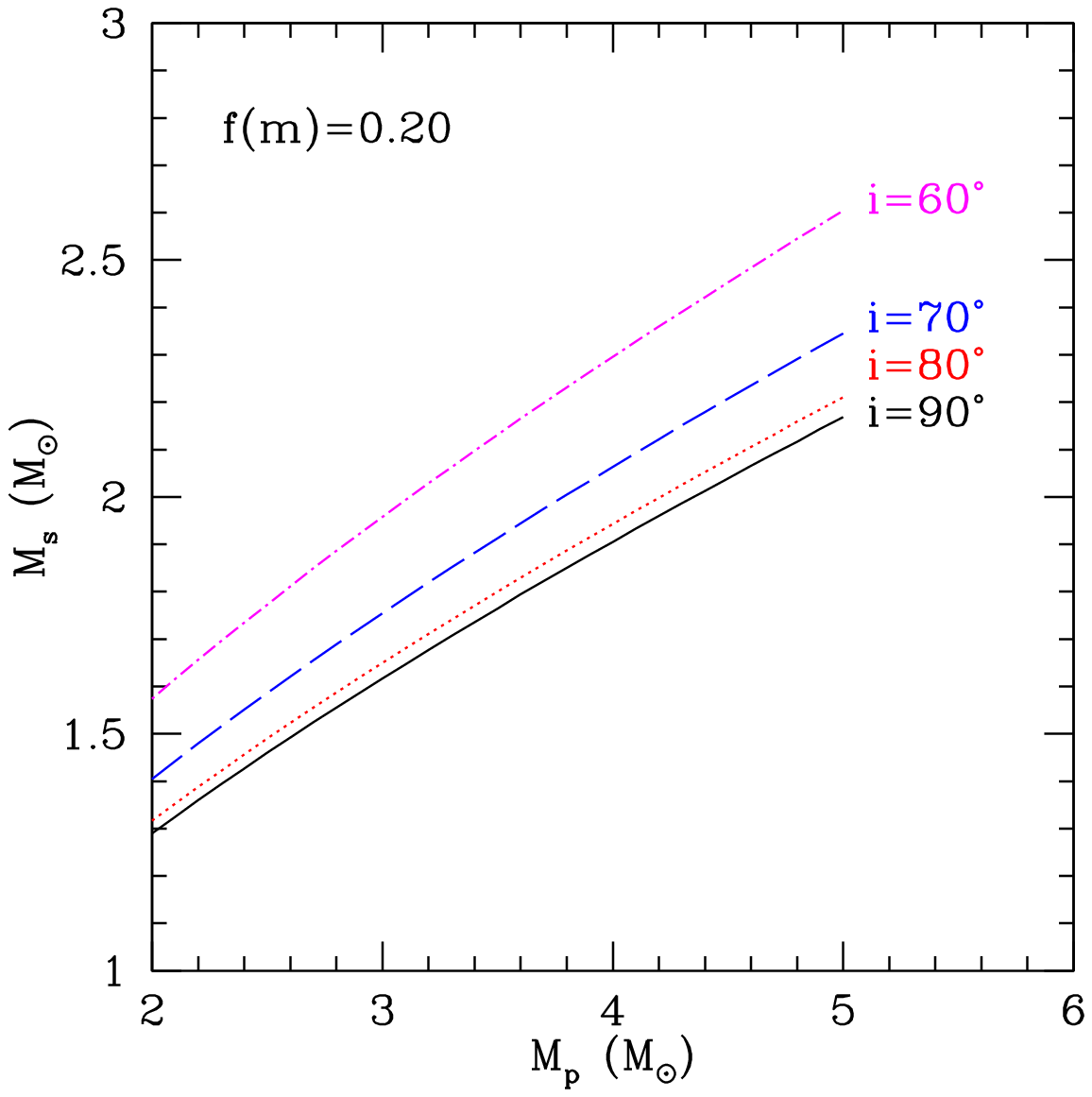}
\caption{\label{fig:mf} Masses for the Undetected Secondary in the 2.2-day Binary.   The mass ($M_s$) of the secondary is shown as a function of the mass ($M_p$) of the primary, the star that is showing He\,{\sc i} $\lambda$4471.  Given the spectral type of the primary is $\approx$ B7~V, we expect its mass to be 3-4$M_\odot$.  Thus the mass of the secondary is likely to be 1.7-2$M_\odot$, which corresponds to a  mid-A through early F-type dwarf.}
\end{figure}

In Figure~\ref{fig:mf} we show the results of the mass function for inclinations from 60$^\circ$ to 90$^\circ$.  Since the system eclipses, we can infer inclination must be high, likely 80$^\circ$ or more.  We see that in the case of either mass function,  the mass of the secondary,  $M_s$, is 1.8-2.7~$M_\odot$. This corresponds to an A2-F0~V star. Such a star would be too cool to contribute to the He\,{\sc i} $\lambda$ 4471 line, although it would contribute to the Balmer lines.  The total mass of the 2.2-day pair would be 5.8-7.7~$M_\odot$. 
 
\begin{figure}[ht]
\epsscale{0.9}
\plotone{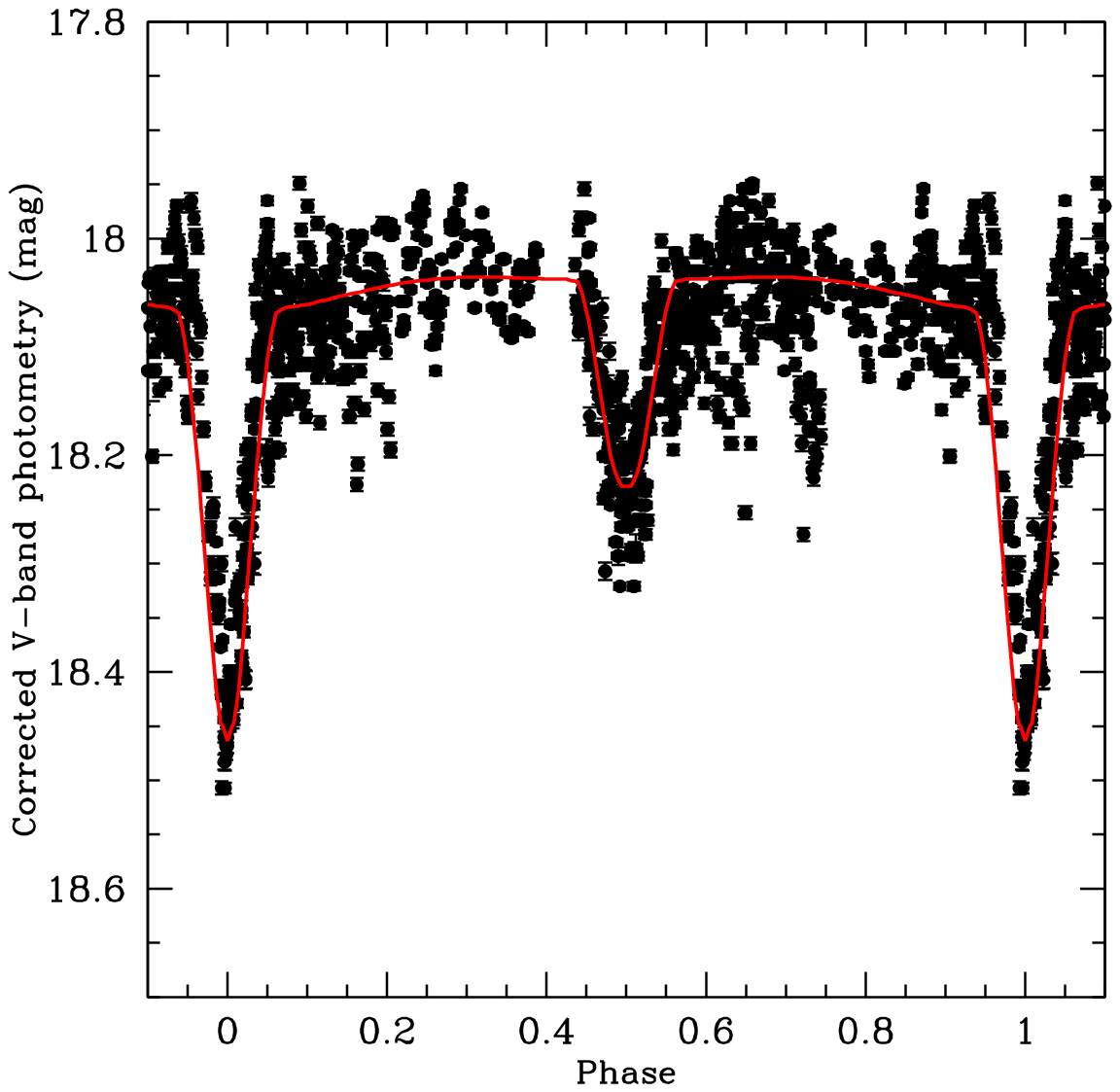}
\caption{\label{fig:lccor} Corrected light-cuve of the 2.2-day period binary.  Here we show the light-curve after subtracting off the expected flux contribution of the WR component, along with our best-fit GENSYN model (red).
}
\end{figure}

With our knowledge of the relative contribution of the WR star to the combined flux, we can now correct the light curve.
We have found from our SED modeling (Figure~\ref{fig:SED}) that the flux of of the WR star $F_{\rm WR}$ must be scaled by a factor of 1.22 to match the total flux of the system, $F_{\rm WR} + F_{\rm 2.2d-binary}$.  Given that the total system  magnitude out of eclipse has $V=16.17$, we subtract $1/1.22 \times 10^{(16.17/-2.5)}$ from the combined flux.  We show the corrected light-curve in Figure~\ref{fig:lccor}.   The eclipses are now no longer shallow.

Modeling this corrected light-curve now allows us to pin down the inclination of the system, as well as further constrain the physical parameters of the system.  To do this, we used the light curve synthesis code GENSYN \citep{1972MNRAS.156...51M}.
Our approach was to make a constrained fit using as much data as possible from the spectroscopic analysis.
We adopted $K_p$ for the 2.2-day primary of 96 km s$^{-1}$, our preferred solution as discussed above.  From the presence of He\,{\sc i} $\lambda$4471 but (apparent) lack of He II absorption, we assume the primary is a B-type object with a temperature range of 12000 – 15000~K. And finally, the combined $M_V$ of the binary is $-0.9$.  

We estimate the physical fluxes and limb darkening coefficients from values from \citet{1979ApJS...40....1K} and \citet{1985A&AS...60..471W}, respectively. Each trial run of GENSYN is set by four independent parameters, the system inclination $i$, the mass ratio of the binary $q$, and the primary and secondary filling factors. GENSYN defines the filling factor to be the ratio of the photospheric surface to the Roche surface. The radii of the two stars are set from the semi-amplitude value of the primary, the mass ratio, and the filling factors.

For each run, we attempt to match three observables: the absolute visual magnitude of the system, the eclipse depths, and the eclipse widths. Our best fit solution is that with the calculated $M_V$ of the system that also matched the eclipse depths and widths. The ratio of eclipse depths is highly dependent upon the temperature difference between the two stars, while the calculated $M_V$ requires the primary’s $T_{\rm eff}$ to be on the higher end of the range estimated from the spectra. The radii and mass ratio are constrained by both the calculated $M_V$ of the system and the eclipse widths. Smaller radii result in both a fainter binary and narrower eclipses, while decreasing the mass ratio produces narrower and shallower eclipses. And finally, the depth of the eclipses constrains the inclination.

Our best solution is presented in Figure~\ref{fig:lccor} and Table~\ref{tab:FinalModel}. All of the parameters agree well with our initial estimates. The inclination is well constrained and exceeds 80 degrees. The secondary appears to be an early A-type dwarf; too cool to contribute to the He\,{\sc i} $\lambda$4471 feature. The visual flux ratio indicates the primary is the main source of the additional optical flux. And the total mass of the system $\sim$6 $M_\odot$. 

We do caution that our modeling and interpretation is premised on the fact that only one component of the LMCe055-1 system contributes significantly to the He\,{\sc i} $\lambda$4471 line.  We do not see this feature as ever doubling, but the weakness of this feature precludes commenting on whether the line shape varies from spectrum to spectrum or not. (This is despite the long exposure times with large aperture telelescopes.)  If the He\,{\sc i} line is a blend, then the actual orbital amplitude of the 2.2-d primary would likely be bigger, and the actual mass of the secondary larger, than that computed here.

\begin{deluxetable}{l c}
\tabletypesize{\scriptsize}
\tablecaption{\label{tab:FinalModel}Derived Properties of the 2.2-day Binary}
\tablewidth{0pt}
\tablehead{
\colhead{Property}
&\colhead{Value}
}
\startdata
$i$ (deg) & $84\pm2$ \\
$q$ &  $0.48\pm0.02$ \\
$T_{\rm eff}$ primary (K) & $15000\pm500$\\
$T_{\rm eff}$ secondary  (K) &  $9500\pm1000$ \\
$M_p$ ($M_\odot$) & $4.00\pm0.40$ \\
$M_s$ ($M_\odot$) & $1.91\pm0.20$ \\
$R_p$ ($R_\odot$)  & $3.20\pm0.06$ \\
$R_s$ ($R_\odot$) & $1.86\pm0.15$ \\
visual flux ratio  &  0.160 \\
\enddata
\end{deluxetable}

\subsection{The WR Star}

If this is truly a physical triple, we would expect to see {\it some} orbital motion of the WR star.  As noted earlier, its radial velocities range from 224 to 265 km~s$^{-1}$, considerably greater than the 1-5 km~s$^{-1}$ measuring uncertainties (Table~\ref{tab:RVs}).     A  Lafler-Kinman  \citep{1965ApJS...11..216L} period search of the WR radial velocities reveals a strong minimum   ($\theta=0.25$) at  34.93 days, providing impetus to find a viable orbit solution.  

We succeeded by using a combination of old and new methods.  It has long been recognized that shorter period binaries tend to have nearly circular orbits, while longer period systems have higher eccentricities (e.g., \citealt{1910PASP...22...47C}), a fact now attributed to tidal effects (see discussion in \citealt{1995A&A...296..709V} and references therein), and it is notoriously difficult to determine orbital parameters from sparse, poorly sampled data for eccentric systems, as is the case here.  Markov Chain Monte Carlo (MCMC) orbit solvers now abound in order to facilitate detection and characterization of extrasolar planets from the relax radial velocity variations of their parent stars, and we utilized the on-line tools provided by the Data Analysis Center for Exoplanets (DACE) at the University of Geneva.\footnote{https://dace.unige.ch/radialVelocities/}  However, we had difficulties achieving convergence without first choosing reasonable priors for the preliminary elements: despite its sophistication,  the Fourier method used by the DACE system \citep{2016A&A...590A.134D} had difficulties with our data, as well as other highly eccentric systems we had previously analyzed, such as HD~166734 \citep{1980ApJ...238..184C} and HD~15558 \citep{1981PASP...93..500G}. Still, with one or more priors set to the published elements, the DACE system had no problem reproducing the other terms, providing what are likely better estimates of the other parameters.  Those published orbits had originally been found using a FORTRAN program described by \citet{1978ApJ...224..558M} based on the classical \citet{1967mamt.book..251W} least-squares, differential-corrections algorithm which utilizes a harmonic analysis for determining the preliminary elements\footnote{We recommend Dr.\ Jeremy B. Tatum's excellent on-line introduction to the analysis of radial velocities of spectroscopic binaries to anyone revisiting the subject; it can be found at https://www.astro.uvic.ca/~tatum/celmechs/celm18.pdf.}.   We resurrected the code to analyze the WR radial velocities here.  

 There is a roughly 20 km~s$^{-1}$ decrease  in radial velocities from 2458153 to 2458154.  If correct, this requires a high eccentricity given the 35 day period.  As tempting as it may be to dismiss one or the others of these measures as inaccurate, they are actually based on several independently calibrated spectra on those two nights. (Many binary star observers have learned the lesson of ignoring a single seemingly anomalous data point.)  
 
\begin{figure}
\epsscale{0.45}
\plotone{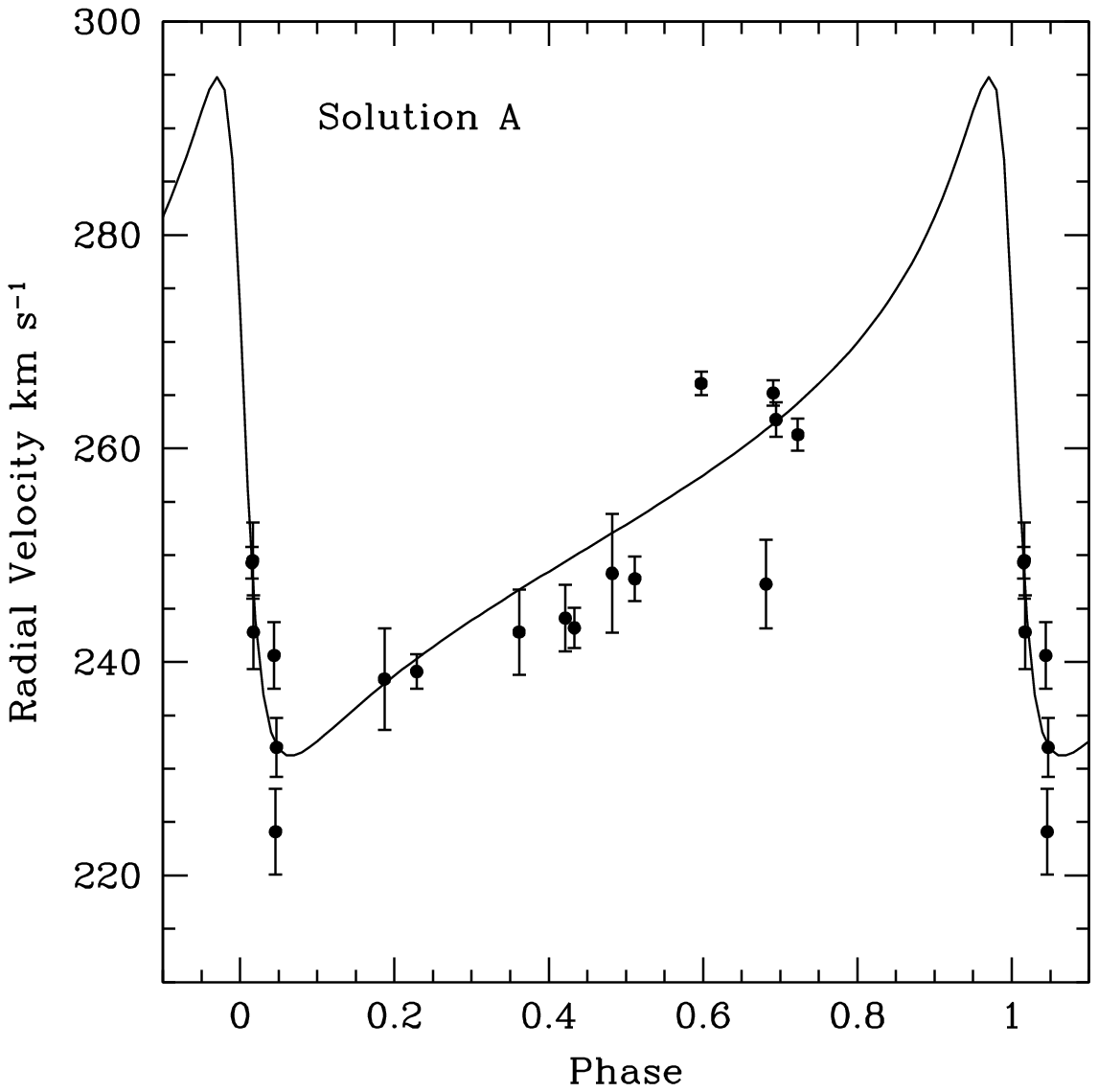}
\plotone{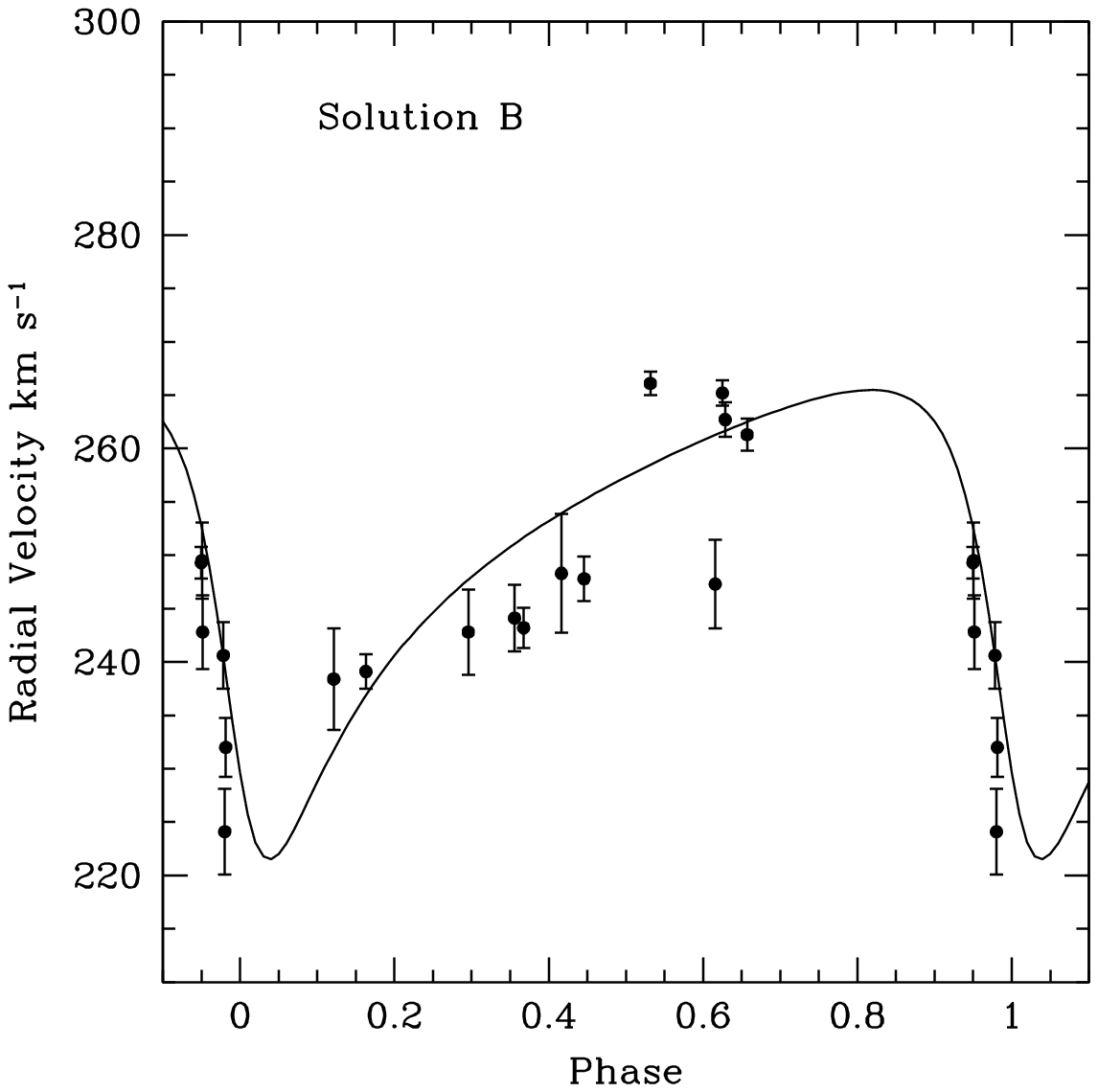}
\caption{\label{fig:orbits} Two Possible Orbit Solutions for the WR Component.  The parameters are given in Table~\ref{tab:orbits}.}
\end{figure}

 Our efforts identified two possible orbit solutions.  We present the parameters in Table~\ref{tab:orbits} and show the comparison with the data in Figure~\ref{fig:orbits}. (Note that the quantity ``R1" is an estimate of the goodness-of-fit of the orbit, calculated from the residuals and the number of degrees of freedom.) We do not consider either of these particularly convincing, much less definite.   Solution A lacks coverage during the critical phases of 0.8 and 1.0; without those data it is hard to place much reliance in it.  Solution B is less eccentric but the curve is not a particularly good representative of the data. 
  \begin{deluxetable}{l c c c}
\tabletypesize{\scriptsize}
\tablecaption{\label{tab:orbits} Possible Orbit Solutions for the WR Component}
\tablewidth{0pt}
\tablehead{
\colhead{Parameter}
&\colhead{Solution A}
&\colhead{Solution B}
}
\startdata
$P$ (days)              & 34.959$\pm$0.002 &  34.959$\pm$0.03  \\
$\gamma$ (km~s$^{-1}$)  &  255.7$\pm$0.6   &   251.1$\pm$1.0    \\
$K$ (km~s$^{-1}$)       &   31.8$\pm$1.0   &   22.0$\pm$2.4    \\
$e$                     &   0.72$\pm$0.02  &   0.55$\pm$0.03   \\
$\omega$ (deg)          &   71.4$\pm$3.0   &  129.0$\pm$15.0   \\
$a \sin i$ (km)             & 1.06E7    &  8.83E6 & \\
$a \sin i$ ($R_\odot$)  & 15.2   & 12.7  \\
$T_0$ (HJD-2450000)     &    9236.7        &   9239.0           \\
$f(m)$ ($M_\odot$)      &      0.039       &   0.023    \\
R1 (km~s$^{-1}$)          &       2.8        &   3.4      \\
\enddata
\end{deluxetable}

\begin{figure}
\epsscale{0.5}
\plotone{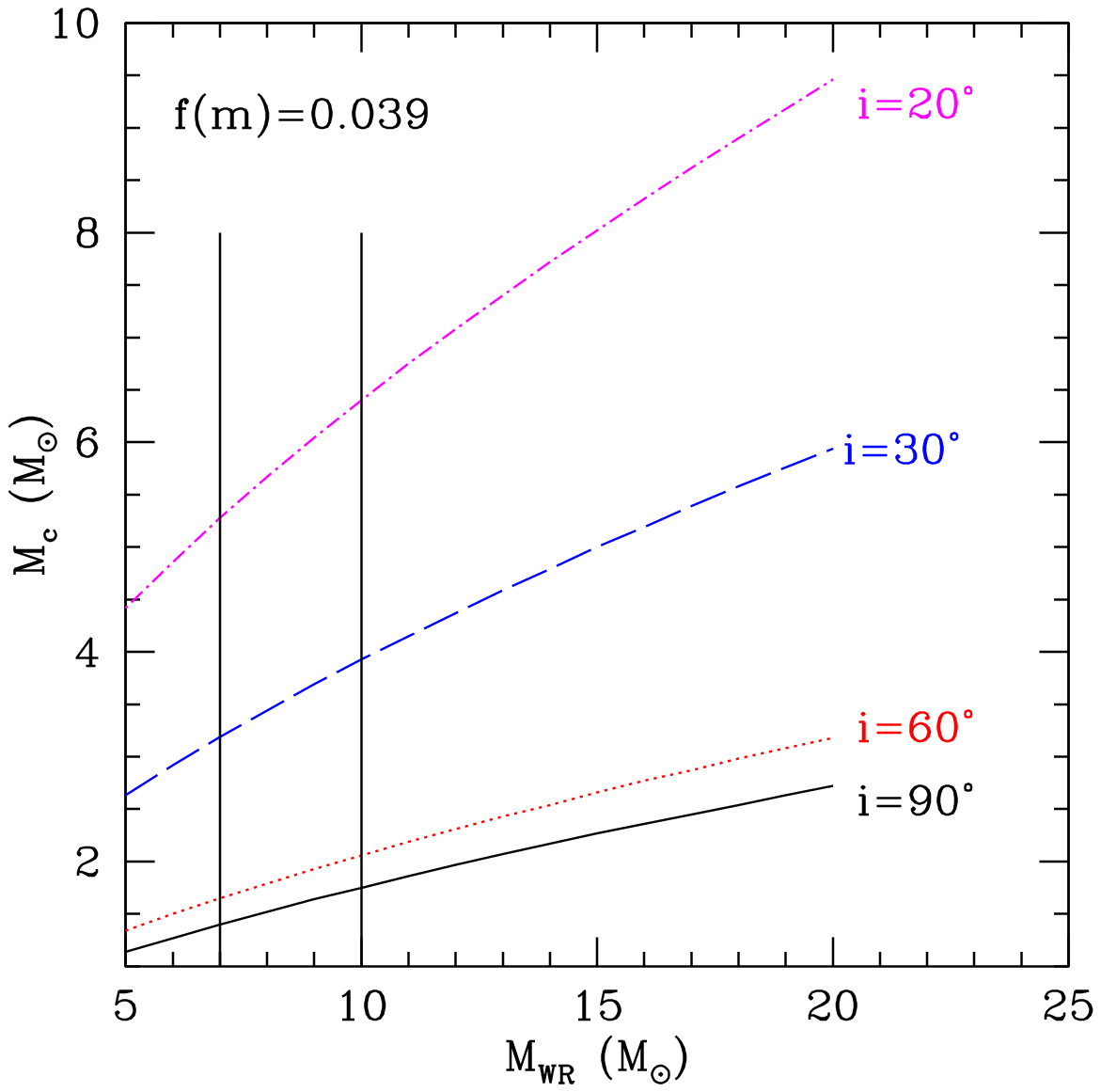}
\caption{\label{fig:MFWR}. Masses for the  secondary in the 35-day WR binary.  Using the mass function from Solution A (Table~\ref{tab:orbits}), we have plotted the mass $M_c$ of the  companion of the WR star as a function of the WR star's mass and orbital inclination.  The two solid vertical lines denote our expected masses for the WR star. Thus for the companion to match the masses of the 2.2-day pair, 5.8-7.7$M_\odot$, would require an orbital inclination of $<$$20^\circ$.}
\end{figure}

The mass functions of the WR orbit solutions are quite small, 0.02 to 0.04 $M_\odot$.   We show the implications of these mass functions in Figure~\ref{fig:MFWR}.   We explore the implications in the following analysis. 

\subsection{Putting it All Together: Three scenarios}

Although we have concerns about the orbit solutions for the WR star, there can be little question that it is showing motion in excess of the measuring uncertainties.  Thus, the WR star has some companion, either the 2.2-d binary or an unseen companion. We see three possibilities that would explain all the data.  We describe these here.

\subsubsection{A Captured Triple}

The simplest explanation would appear to be that the system is a triple, with the 2.2-day pair in a 35-day period orbit
around the WR star.  However, there are complications with this scenario.  First, 
 if our analysis of the 2.2-day binary is correct, we expect it to have a total mass of $\sim$6$M_\odot$.  Even if the WR star's mass was as large as 20$M_\odot$, the inclination would have to be $<$$30^\circ$ according to the mass functions shown in Figure~\ref{fig:MFWR}. Thus, if this were a physical triple system, the three stars would not be orbiting coplanarly.  This would pose challenges to formation mechanisms, let alone dynamical stability issues, although such systems are said to exist (e.g., Herschel 36, \citealt{2019MNRAS.484.2137C}).

 We note that most triple systems are described as hierarchical, with the close pair more massive and the third, more distant body, being of lower mass.   A triple consisting of the WR star as the primary and the 2.2-day binary as the secondary would run contrary to this,  but such systems are known.  One of them, HD 181068,  even contains planets! (See discussion in \citealt{2018A&A...619A..91B}.)  However, that system is coplanar, with all three stars mutually eclipsing. The low amplitude found here for the WR star's motion does not allow it to be coplanar with the orbital motion within the 2.2 day binary.
 
 An even more serious concern, however, involves the evolutionary ages.  We do not know the initial mass of the WR component, but it is reasonable to suppose that it is initially of 30$M_\odot$ or more.  Massive stars form quickly (10$^4$-10$^{5}$ yr), and burn
through their hydrogen in a few million years. The age of a WR star is typically taken to be 3-5 Myr, at least in single-star evolutionary models (see, e.g., Table 1 in \citealt{2012A&A...542A..29G}).  We are concerned by the formation time of the
2.2-day pair.  The B7~V primary with a mass of 4$M_\odot$ does not pose much of a problem, as its contraction time would be of
order 5 Myr according to the Z=0.02 models of \citet{1996A&A...307..829B}. However, the 1.9$M_\odot$ secondary is more problematic, as its contraction time is $\sim$25~Myr.   Caught only 5~Myr into its formation such a pre-main-sequence star would have a huge radius, a possibility quite ruled out by the light curve.

If the WR star and the 2.2-day pair form a triple system, we see only one possible explanation that would encompas all of these weird aspects: perhaps this is a case of a captured triple, with the more massive star ``capturing" the (older) 2.2-day pair at some earlier point in its life.  Such a scenario would explain (a) the lack of coplanarity, (b) the different inferred ages of the WR star and the 2.2-day pair,  (c) the high eccentricity of the 35-day orbit, and (d) the fact that this system runs counter to the usual hierarchical structure.
Such capture scenarios have been shown to be inefficient for solar-mass stars, but are much more efficient for massive
capturers, as shown by the modeling simulations of \citet{2007ApJ...656..275M}.

\subsubsection{A Physical Quadruple System}
A more likely possibility would be that the WR star has an unseen companion in a 35 day orbit, and that the 2.2-day pair is in a longer period orbit about the pair.   If we drop the requirement of coplanarity then the mass of the secondary could be high enough
to not pose an age problem.  For instance, if the inclination is 30$^\circ$, the companion could have a mass of 3-4$M_\odot$, similar to that of the primary of the 2.2-day pair.  However such a star would certainly be hot enough to have He\,{\sc i} $\lambda$4471.  We saw no sign of the He\,{\sc i} being double in any of our data, but give how weak the line is, we cannot exclude this.  It would certainly affect our orbit determination of the 2.2-d primary, likely reducing the orbital amplitude.  If the orbital semi-amplitude of the primary was larger, with the same period and orbital inclination, 
the mass of the secondary of the 2.2-d pair would also be larger, eliminating the age problem and no longer  requiring a capture explanation for  the two pairs.  We do not dislike this scenario, but note that it is more {\it ad hoc}, requiring yet another unseen companion.

\subsubsection{A Line of Sight Quadruple System}
 
 Might the two pairs of stars be a line of sight coincidence; i.e., rather than the system being a triple, might we just be viewing two binary pairs superimposed on the sky? Gaia ER3 detects no other star within 2\arcsec\ of LMCe055-1, and its ``renormalized unit weight error" (RUWE) is 1.08, consistent with there being no indication of multiplicity \citep{2021A&A...649A...2L}. Given the magnitude difference, however,  this corresponds to a spatial resolution of about 1\arcsec\ \citep{2021A&A...649A...6G}.\footnote{The interpretation of RUWE is nicely discussed in an unpublished 
document by Logan Pearce, available at http://www.loganpearcescience.com/research/RUWE\_as\_an\_indicator\_of\_multiplicity.pdf.}  

\citet{LMCBins} identified the percentage of red supergiant (RSG) binaries in  the LMC.  Of the systems found to be binaries, all had B-type companions.  They describe a Monte Carlo simulation they performed in order to address what fraction of the RSG+B systems might be line of sight coincidences, with the B star falling into the 1\arcsec\ wide observing slit.   
We followed their methodology here: we first calculated the surface density of OB stars within 5\arcmin\ of LMCe055-1 from the photometry of \citet{ZaritskyLMC}, using as our criteria $V\leq 19.5$ and $B-V<0.13$, roughly corresponding to an A0~V star.  There are 1297 such stars (not counting LMCe055-1 itself), leading to a surface density of 4.59 $\times 10^{-3}$ stars arcsec$^{-2}$.  We then ran 100,000 Monte Carlo simulations, assuming a Poisson distribution of the number stars, and randomizing their spatial distribution.   The probability of any star landing within 1\arcsec\ of a particular position is 0.4\%.  In practice this is probably an overestimate, as many of our spectroscopic observations were made in sub-arcsecond seeing and the slit had a half-radius of 0\farcs5, but the value is illustrative: it is highly unlikely that this is a chance line-of-sight.

\subsubsection{But has the WR star been stripped?}

The major motivation for this study has been to determine if LMCe055-1 is an example---and possibly an archetype--- of a stripped WR binary.  Our analysis effectively ruled this out.  The distance of closest approach (pericenter) $r_p$ to the center of mass of the system will just be $r_p\sin i = a \sin i (1-e)$, or about 3.0E6 km (4.0$R_\odot$) or 4.0E6 km (5.7$R_\odot$), using Solutions A or B, respectively.  This is much larger than the star's radius (2.4$R_\odot$, from Table~4).  Of course, as the WR progenitor  lost mass, the orbit itself would evolve, and the system could have been closer at some point. Similarly, if the WR component had evolved to cooler temperatures (larger radii) there could also have been interactions in the past.  However, the relatively small masses of the 2.2-d pair show that little mass accretion could have taken place in the past.  We therefore believe that the answer to the question posed in the title is a likely ``no."


\section{Summary and Conclusions}
\label{Sec-conclude}

LMCe055-1 was originally classified as a WN4/O4, similar to the nine known LMC WN3/O3 stars, but of somewhat lower excitation given the relative strength of N\,{\sc iv} and N\,{\sc v}, and the presence of weak He\,{\sc i} $\lambda$4471 absorption in addition to the rich H and  He\,{\sc ii} absorption spectrum.   The star shows shallow eclipses, with a 2.2 day period.  Our study here has shown that the He\,{\sc i} absorption moves in concert with the eclipse period, but in such a way that it cannot be the secondary of the WR star but must instead be the primary of the 2.2-day pair.   Removing the flux of the WR component from the light-curve allowed us to determine the physical properties of the components of the 2.2-day period system in concert with our radial velocity study. We find the 2.2 d system consists of a 4.0$M_\odot$ and 1.9$M_\odot$ pair.  

We have explored the relationship of this pair to that of the WR star.  The WR radial velocities show variations of order 30-40 km~s$^{-1}$, greater than our measuring uncertainties.  Our best fit shows a 35-day eccentric orbit.  If the ``secondary" of this motion is the 2.2-d binary pair, then the triple system is not coplanar.  The contraction time of the secondary of the 2.2-day pair is 25~Myr, which we expect is much greater than the lifetime of the WR star.  Such a triple system would be neither coplanar nor coeval.  The only way we can envision such a system to form would be that  the WR star ``captured" the 2.2-day pair earlier in its life.   This would explain both the lack of coplanarity, the high eccentricity of the 35-day orbit, the mismatch in stellar ages, and the problem that this system runs counter to the usual hierarchical structure.

A more promising possibility is that the system is a physical quadruple.  The WR star would have an unseen companion in a 35 day eccentric orbit.  If the orbital inclination is 30$^\circ$, the companion would be of a mass comparable to that of the primary in the 2.2-day system.  Such a star would have He\,{\sc i} $\lambda$4471 absorption, and would have decreased the actual orbital amplitude we measure. This would mean that the secondary of the 2.2-day system would have a larger mass, and remove the issue of the lack of coevality. It would still require the system to not be coplanar.

This multiplicity does not make LMCe055-1 unique, or even unusual. Triple and even quadruple star systems may be fairly common among early-type stars, unlike those of solar-type stars and later \citep{2017ApJS..230...15M}. A well known example is the WR system HD 211853 \citep{1981ApJ...244..157M}.  A similar model has been proposed for [M2002] LMC 172231 \citep{2024ApJ...967...64T}.  Possibly the most famous example of all, however, is HD 5980,
a complex system that is likely quadruple \citep{2014AJ....148...62K,2019MNRAS.486..725H,2022RMxAA..58..403K}.

What does make LMCe055-1 singularly interesting is its spectral characteristics and derived physical properties.
The nine LMC WN3/O3 stars all have similar physical properties to one another \citep{NeugentWN3O3s}.  A comparison with LMCe055-1 shows important differences: LMCe055-1 is cooler and has an even lower mass-loss rate. It is likely less massive.  The He/H number ratio is 2, which might suggest it is slightly more evolved, but recall that the value is highly dependent on how we compensate for the B star component: before correction the value was 0.15.
 
We began this study in the hopes of resolving whether or not LMCe055-1 is a stripped binary.  Had the WR component been a member of a pair in a 2.2-day orbit, then it most certainly would have undergone massive stripping during its short lifetime.    However, we have shown that the short period and eclipses have nothing to do with the WR star. 

\citet{NeugentWN3O3s} proposed that the WN3/O3 stars represent a short-lived stage in the normal evolution of massive stars.  The widths of their absorption lines suggest projected rotational velocities of 100-120 km~s$^{-1}$, quite unremarkable for massive stars, and suggestive that these stars have not undergone mergers.  We find the same is true for LMCe055-1.  If our model of the LMCe055-1 system is correct, then its current multiplicity provides a further argument against the WR star having formed by stripping: the current members of the system do not come close enough to interact,
and the low mass of the 2.2-d pair argues that no significant amount of mass was accreted in the past. 

\begin{acknowledgements}

Lowell Observatory sits at the base of mountains sacred to tribes throughout the region. We honor their past, present, and future generations, who have lived here for millennia and will forever call this place home.   The observations presented here were obtained over years at Las Campanas Observatory, and we are grateful to the excellent technical and logistical support we have always received there. We also acknowledge long-term support by both the Carnegie and Arizona Time Allocation Committees.

We thank the anonymous referee for useful comments which improved the presentation. We are grateful to Drs.\ Konstantina Boutsia, Carlos Contreras, and Maria Drout for obtaining MagE observations at critical phases.  We also appreciate the expert support provided by the Gemini-S GMOS queue observer Pablo Prado, as well as the assistance in reducing those data by Dr.\ Vinicius Placco.   Dr.\ George Jacoby kindly spent several hours obtaining photometry with the Swope in our efforts to improve coverage of the eclipses. We are also indebted to Dr.\ Alex Fullerton for a nice comparison of the fluxes of the various COS UV spectra of LMCe055-1, as well as an account of how our star wound up on the ULLYSES observing list.  We also appreciate a very useful conversation over coffee with Dr.\ Noel Richardson, which helped lead to the conclusion that the system was not just a binary but a multiple.  We received useful guidance from Drs.\ Joe Llama and John Michael Brewer for using DACE, and we acknowledge useful correspondence with Dr.\ Philip Bennett on the use of an alternative orbit solver. Dr.\ Deidre Hunter provided a critical reading. The Monte Carlo Python code used in the line-of-sight calculation was written by the ChatGPT-3.5 bot. We thank these future robotic overlords for their assistance.

Partial support for this work was provided by the National Science Foundation through AST-2307594 awarded to P.M and D.J.H.   Additional support for the analysis was provided by NASA through grants GO-6299.001 and AR-17553.001.

This work has made use of data from the European Space Agency (ESA) mission
{\it Gaia} (\url{https://www.cosmos.esa.int/gaia}), processed by the {\it Gaia}
Data Processing and Analysis Consortium (DPAC,
\url{https://www.cosmos.esa.int/web/gaia/dpac/consortium}). Funding for the DPAC
has been provided by national institutions, in particular the institutions
participating in the {\it Gaia} Multilateral Agreement.

This work also made use of observations obtained under program GS-2021A-FT-103 at the international Gemini Observatory, a program of NSF’s NOIRLab, which is managed by the Association of Universities for Research in Astronomy (AURA) under a cooperative agreement with the National Science Foundation on behalf of the Gemini Observatory partnership: the National Science Foundation (United States), National Research Council (Canada), Agencia Nacional de Investigaci\'{o}n y Desarrollo (Chile), Ministerio de Ciencia, Tecnolog\'{i}a e Innovaci\'{o}n (Argentina), Minist\'{e}rio da Ci\^{e}ncia, Tecnologia, Inova\c{c}\~{o}es e Comunica\c{c}\~{o}es (Brazil), and Korea Astronomy and Space Science Institute (Republic of Korea).

\end{acknowledgements}
\facilities{Magellan:Baade (MagE), Swope (SITe No. 3 imaging CCD), HST (COS), Gaia}

\bibliographystyle{aasjournal}
\bibliography{masterbib.bib}

\end{document}